\documentclass[aps,prl,superscriptaddress, floatfix, twocolumn]{revtex4-1}

\usepackage{amsmath}
\usepackage{amssymb}
\usepackage{wasysym}
\usepackage{graphicx}
\usepackage{color}
\usepackage{hyperref}
\usepackage{physics}
\usepackage{siunitx}
\usepackage[english,nomargin,inline,marginclue,draft]{fixme}
\fxusetheme{colorsig}
\FXRegisterAuthor{cg}{acg}{CG}  
\makeatletter
\renewcommand*\FXLayoutInline[3]{%
  {\@fxuseface{inline}\ignorespaces{\color{fx#1}[#3: #2]}}}
\makeatother

\begin{document}

\title{Direct observation of incommensurate magnetism in Hubbard chains}

\author{Guillaume~Salomon}
\email{guillaume.salomon@mpq.mpg.de}
\author{Joannis~Koepsell}
\author{Jayadev~Vijayan}
\author{Timon~A.~Hilker}
\affiliation{Max-Planck-Institut f\"{u}r Quantenoptik, 85748 Garching, Germany}
\author{Jacopo~Nespolo}
\affiliation{Fakult\"{a}t f\"{u}r Physik, Ludwig-Maximilians-Universit\"{a}t, 80799 M\"{u}nchen, Germany}
\affiliation{INO-CNR BEC Center and Dipartimento di Fisica, Universita di Trento, 38123 Povo, Italy}
\author{Lode~Pollet}
\affiliation{Fakult\"{a}t f\"{u}r Physik, Ludwig-Maximilians-Universit\"{a}t, 80799 M\"{u}nchen, Germany}
\author{Immanuel~Bloch}
\affiliation{Max-Planck-Institut f\"{u}r Quantenoptik, 85748 Garching, Germany}
\affiliation{Fakult\"{a}t f\"{u}r Physik, Ludwig-Maximilians-Universit\"{a}t, 80799 M\"{u}nchen, Germany}
\author{Christian~Gross}
\affiliation{Max-Planck-Institut f\"{u}r Quantenoptik, 85748 Garching, Germany}

\date{\today}

\begin{abstract}
The interplay between magnetism and doping is at the origin of exotic strongly
correlated electronic phases and can lead to novel forms of magnetic ordering.
One example is the emergence of incommensurate spin-density waves with a wave
vector that does not match the reciprocal lattice. In one dimension this effect
is a hallmark of Luttinger liquid theory, which also describes the low energy physics of
the Hubbard model~\cite{giamarchi2004}. Here we use a quantum simulator based on ultracold fermions
in an optical
lattice~\cite{greif2013,hart2015,parsons2016,boll2016,cheuk2016,drewes2017,brown2017}
to directly observe such incommensurate spin correlations in doped and
spin-imbalanced Hubbard chains using fully spin and density resolved quantum
gas microscopy. Doping is found to induce a linear change of the spin-density
wave vector in excellent agreement with Luttinger theory predictions. For
non-zero polarization we observe a decrease of the wave vector with
magnetization as expected from the Heisenberg model in a magnetic field. We
trace the microscopic origin of these incommensurate correlations to holes,
doublons and excess spins which act as delocalized domain walls for the
antiferromagnetic order. Finally, when inducing interchain coupling we observe
fundamentally different spin correlations around doublons indicating
the formation of a magnetic polaron~\cite{dagotto1994}.\\
\end{abstract}

\maketitle

One dimensional (1D) quantum systems are paradigmatic examples of the breakdown
of Landau-Fermi liquid theory. The free quasiparticle concept present in higher
dimensions is replaced by collective excitations leading to striking phenomena
such as spin-charge separation~\cite{giamarchi2004}. Luttinger liquid
theory~\cite{haldane1981a} generically describes the low energy physics of
gapless one-dimensional systems ranging from quasi-1D conductors, spin liquids
to chiral edge modes in the fractional quantum Hall effect~\cite{wen90}. In
particular, the repulsive single-band Hubbard model, which provides a minimal
microscopic description of doped antiferromagnets, can be described through
this approach. Away from half filling, Luttinger liquid theory predicts
incommensurate magnetism with an algebraically decaying incommensurate
spin-density wave (SDW) at zero temperature, whose vector varies linearly with
density~\cite{giamarchi2004}. Also, the presence of a spin imbalance in the 1D
Hubbard model can lead to incommensurate spin
correlations~\cite{frahm1991}. Short-range incommensurate magnetism is expected
to survive at finite temperature, where conformal field theory arguments
predict an exponential decay of the spin correlations with
distance~\cite{cardy1996}. Luttinger liquids were experimentally studied in
traditional condensed matter systems such as carbon nanotubes via conductance
and scanning tunneling microscopy measurements~\cite{bockrath1999,lee2004}, and
in particular, magnetism was studied through neutron scattering on weakly
coupled quasi-1D spin-1/2 chains~\cite{stone2003,lake2005} and on ladder
systems~\cite{klanjsek2008}. In higher dimensions, incommensurate spin-density
waves were detected in the underdoped region of certain high-T$_\mathrm{c}$
superconductors via neutron scattering~\cite{tranquada1995}. An interpretation
in terms of holes organized in stripes was proposed, which results in an effective 1D description, where the stripes form domain walls in the
antiferromagnet. Here we use real space spin and density resolved quantum
gas microscopy to directly study the effects of both doping and polarization on
finite range spin correlations in the 1D Hubbard model. We measure the linear
change in the SDW vector as a function of density in excellent agreement with
Quantum Monte-Carlo (QMC) calculations. In presence of a population imbalance,
we observe an increase of the SDW wavelength with polarization as predicted by
Luttinger liquid theory. We finally report on the evolution of the
antiferromagnetic spin correlations around doublons in the crossover from 1D to 2D. 
We find the magnetic environment around doublons to change
fundamentally when spin correlations appear in the transverse direction,
suggesting the formation of a magnetic polaron~\cite{dagotto1994}.

\begin{figure*}[t]
\centering
\includegraphics{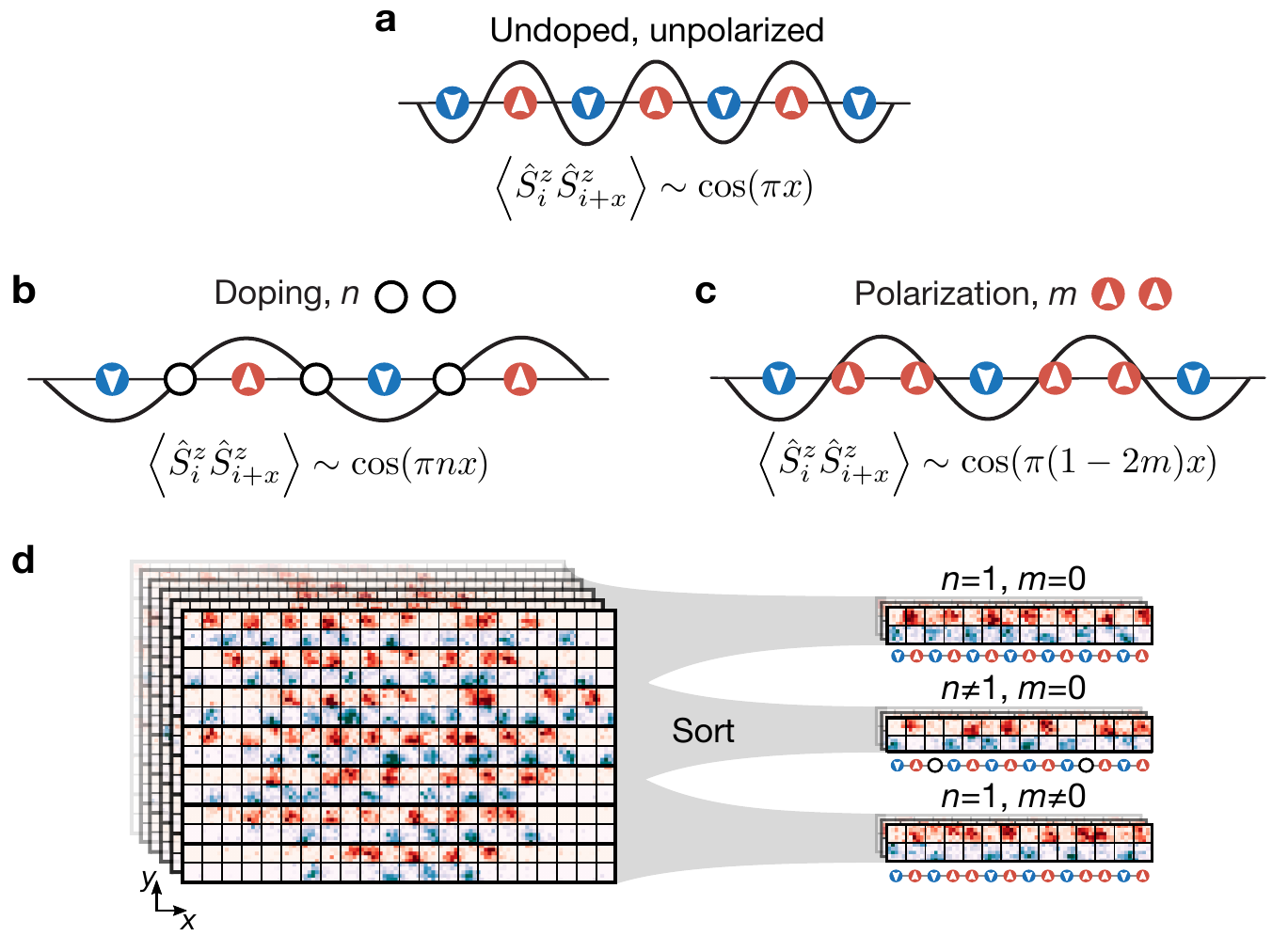}
\caption{\textbf{Probing incommensurate spin correlations in Hubbard chains.}
\textbf{a}, Spin correlations in spin-balanced Hubbard chains at half filling
($n=1$) form at a commensurate wave vector $\pi$. \textbf{b}, When the system
is doped ($n\neq1$), incommensurate spin correlations at a wave vector $\pi n$
develop due to delocalized holes and doublons, which act as quantum domain walls
stretching the distance between antiferromagnetically correlated spins.
\textbf{c}, At finite polarization $m\neq0$, incommensurate spin correlations at
a wave vector $\pi (1-2m)$ arise due to excess spins. \textbf{d}, Left:
Single-spin and density-resolved experimental images, each containing  $7$ independent
Hubbard chains along $y$ separated by thick lines where spins $\uparrow$
($\downarrow$) are represented in red (blue). Right: In post-analysis we group
the data by magnetization and doping to analyze their individual effect on spin
correlations along $x$.}
\label{fig:cartoon}
\end{figure*}

Our experiments started by loading a balanced two-dimensional degenerate spin
mixture of $^6$Li atoms in the lowest two Zeeman states
$\ket{\uparrow},\ket{\downarrow}$ into an optical lattice formed by two
standing waves with period $d_x=1.15\,\mu$m in $x$ direction and
$d_y=2.3\,\mu$m in $y$ direction (Fig.~\ref{fig:cartoon})~\cite{boll2016}. The atoms were
trapped in a single plane of a vertical lattice with $3.1\,\mu$m spacing and a
depth of $17\,E^z_r$ where $E^i_r$ denotes the recoil energy in direction $i$.
The nearest-neighbor tunneling rates were set to $t_x=h\times410\,$Hz at
5$\,E^x_r$ lattice depth and $t_y=h\times 1.2\,$Hz at 27$\,E^y_r$ to study the
one dimensional Hubbard model. By decreasing the lattice depth in the $y$
direction and ramping up the $x$ lattice power to vary $t_y/t_x$, we can
explore the Hubbard model from 1D to 2D. The onsite interaction $U$ was
controlled using the broad Feshbach resonance located at 834.1G and set to
$U=7\,t_x$ in the 1D regime. We directly measured the occupation and spin on
each lattice site by first freezing the atomic motion before a local
Stern-Gerlach like splitting of the spin components in a superlattice along $y$~\cite{boll2016} (Fig.~\ref{fig:cartoon}). Finally we detected the atoms via Raman sideband cooling~\cite{omran2015}. Thanks to the ultimate
resolution  of our detection down to single atoms and spins, we are able to
group our data according to total spin $S^z=(N_\uparrow -N_\downarrow)/2$ and
total atom number $N=N_\uparrow +N_\downarrow$, that is, the sum of the up and down
spin number in each chain. These conserved quantities fluctuate for different
chains and experimental runs (see Supplementary Information), however, data grouping
allows us to individually explore the effect of doping and spin imbalance
(Fig.~\ref{fig:cartoon}).

\begin{figure*}[t]
\centering
\includegraphics{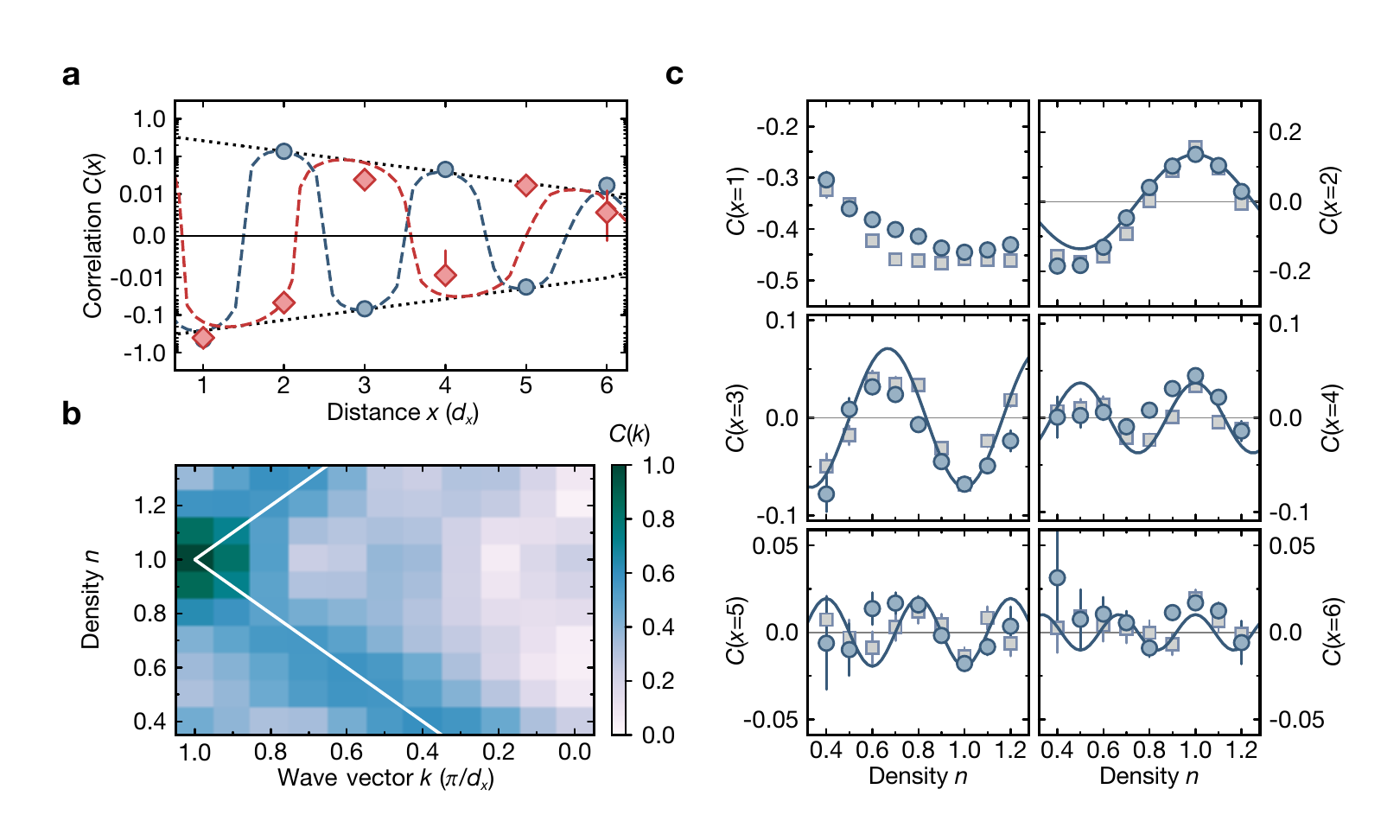}
\caption{\textbf{Incommensurate spin correlations vs doping.} \textbf{a,} Spin
correlations $C(x)$ at half filling (blue) and at $n=0.7$ (red). The dotted
lines show the decay obtained from an exponential fit of the rectified spin
correlations $(-1)^x C(x)$ at half filling. The dashed lines are the Luttinger
liquid predictions using the amplitude and decay length obtained from a fit of
the $n=1$ experimental data. The sign change observed for $d\geq2$ in the
doped case originates from delocalized holes acting as quantum domain walls,
which stretch the distance between antiferromagnetically correlated spins.
\textbf{b,} Away from half filling the normalized Fourier transform of the spin
correlations $C(k)$ reveals a linear increase of the SDW wave vector with
density. The white line is the Luttinger liquid result $k_{\mathrm{SDW}} = \pi n$. \textbf{c,}
Spin correlations $C(x)$ vs density at fixed distances $x=1,...,6$ (blue dots)
compared to QMC calculations at $T=0.29\,t_x$ (gray squares). The measured
densities are binned in intervals of $0.1$. The blue lines are the Luttinger
liquid prediction with wave vectors $\pi n$ using the amplitude and decay
length extracted from the fit in a. Around unity filling oscillations up to
distances of $x=6$ are visible as a function of density in agreement with
Luttinger liquid theory. Errorbars denote standard error of the mean.}
\label{fig:sdw_vs_n}
\end{figure*}

We first study the evolution of antiferromagnetic spin correlations along 1D
chains as a function of doping. The correlations are quantified by the
two-point correlation function $C(x)=4\langle S^z_i S^z_{i+x}
\rangle_{\scalebox{0.65}{\newmoon}_i\scalebox{0.65}{\newmoon}_{i+x}}$
conditioned on the sites $i,i+x$ being singly occupied (filled circles).
Experimentally, we prepared Hubbard chains with up to $N=23$ atoms and
post-selected the experimental outcomes to the $S^z=0$ sector to first consider
the effects of doping only. Due to the underlying harmonic confinement, the
atomic cloud is inhomogeneous and in the spirit of a local density
approximation we define the density $n$ as the mean occupation calculated over
the sites connecting $i$ to $i+x$ for each value of $N$ (see Supplementary
Information). From Luttinger liquid theory one expects the wave vector of the
SDW to be $k_{\mathrm{SDW}}=2k_{\mathrm{F}}=\pi n$ defining the Fermi wave
vector $k_{\mathrm{F}}$. At finite temperature and large distances $x \gtrsim
k^{-1}_{\mathrm{F}}$, the spin correlations are predicted to decay
exponentially~\cite{giamarchi2004}: \begin{equation} C(x) \simeq A\,
e^{-\frac{x}{\xi}} \cos({\pi n x}), \label{density_eq} \end{equation} where $A$
is a non-universal constant and $\xi$ is the temperature-dependent correlation
length that weakly varies~\cite{giamarchi2004} with density at $U/t=7$. We
determined $A$ and $\xi$ from an exponential fit of $C(x)$ at half filling
($n=1$) for $x=2,..,6$ yielding $A=0.49(4),\,\xi=1.6(1)$
(Fig.~\ref{fig:sdw_vs_n}a) where all distances are expressed in units of the
lattice constant $d_x$. Away from half filling, we observe a linear increase of
the SDW vector both for hole and charge doping as revealed by a Fourier
transform of the rescaled spin correlation  $C(k) =
\mathcal{F}\{A^{-1}e^{x/\xi} C(x)\}$. For a quantitative comparison with
theory, we show in Fig.~\ref{fig:sdw_vs_n}c the spin correlations $C(x)$ as a
function of density $n$ together with QMC calculations for a homogeneous system
at temperature $T=0.29\,t_x$ and the long distance Luttinger prediction of
Eq.~\ref{density_eq}. The spin correlations oscillate with a periodicity
$k_{\mathrm{SDW}}=\pi n$ as expected from Luttinger theory. We attribute the
microscopic origin of the incommensurate correlations to delocalized doublons
and holes, increasing the distance between antiferromagnetically correlated
spins~\cite{hilker2017,kruis2004} and thus, the wavelength of the SDW.

\begin{figure*}[t]
\centering
\includegraphics{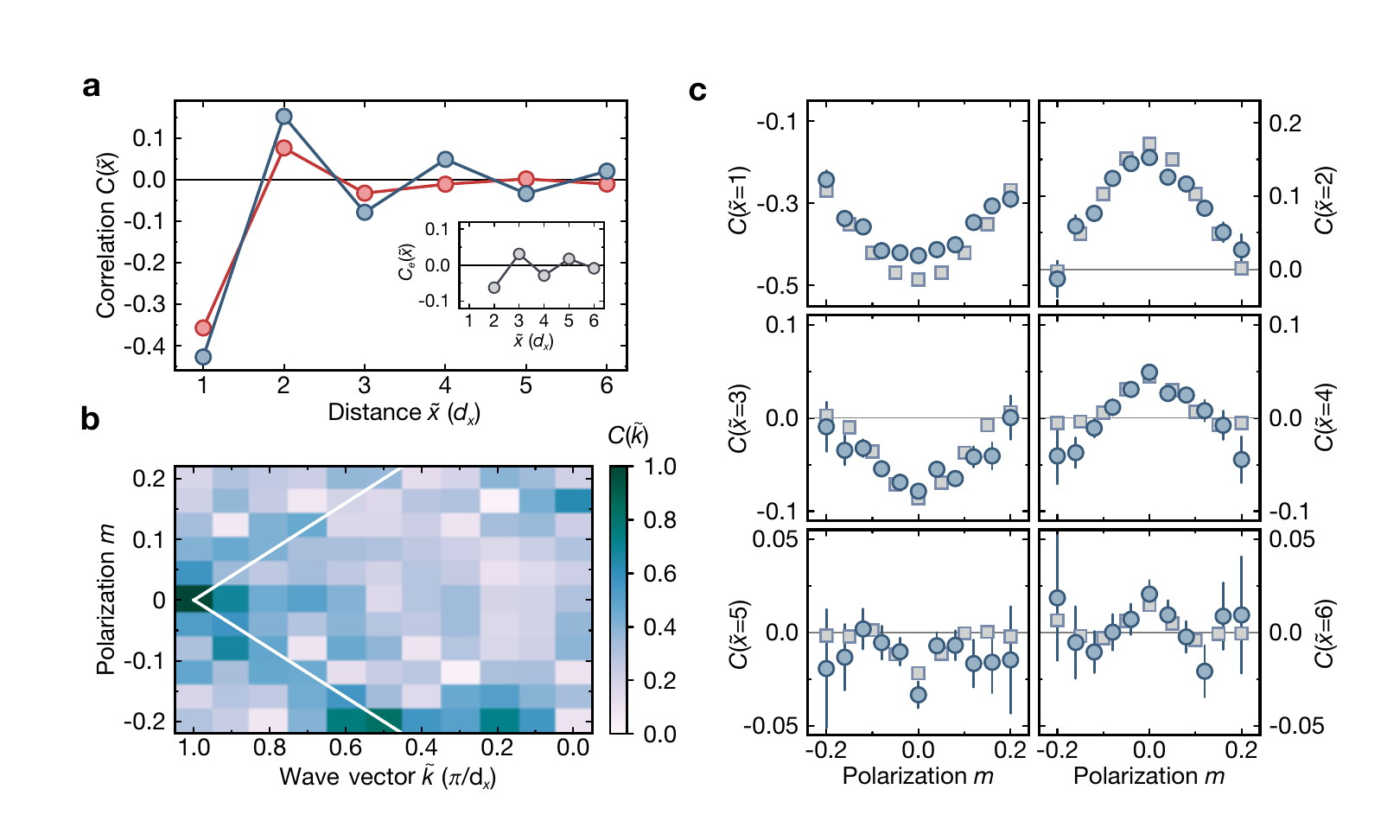}
\caption{\textbf{Incommensurate spin correlations vs polarization.} \textbf{a,}
Spin correlations in squeezed space $C(\tilde{x})$ at $m=0$ (blue) and
$m=-0.12$ (red). A sign change is visible at distance $d>4$ reflecting an
increase of the SDW wavelength away from $m=0$. Inset: In a spin-imbalanced gas at $m=-0.12$
the distance between antiferromagnetically correlated spins is stretched as
revealed by the sign changes in the spin correlations $C_e$ across majority spins
(black).  \textbf{b,} Normalized Fourier transform $C(\tilde{k})$ of
$C(\tilde{x})$ qualitatively revealing a linear change of the SDW vector in
agreement with the Luttinger liquid prediction $\pi (1-2m)$ (white line).
\textbf{c,} Experimental spin correlations $C(\tilde{x})$ vs magnetization $m$
at fixed distances $\tilde{x}=1,...,6$ (blue dots) compared to QMC calculations
at $T/t_x=0.25$ and half filling (gray squares). Binning of the magnetization
is in intervals of $0.04$ and errorbars denote one standard error of the mean.}
\label{fig:sdw_vs_m}
\end{figure*}

\begin{figure*}[t]
\centering
\includegraphics{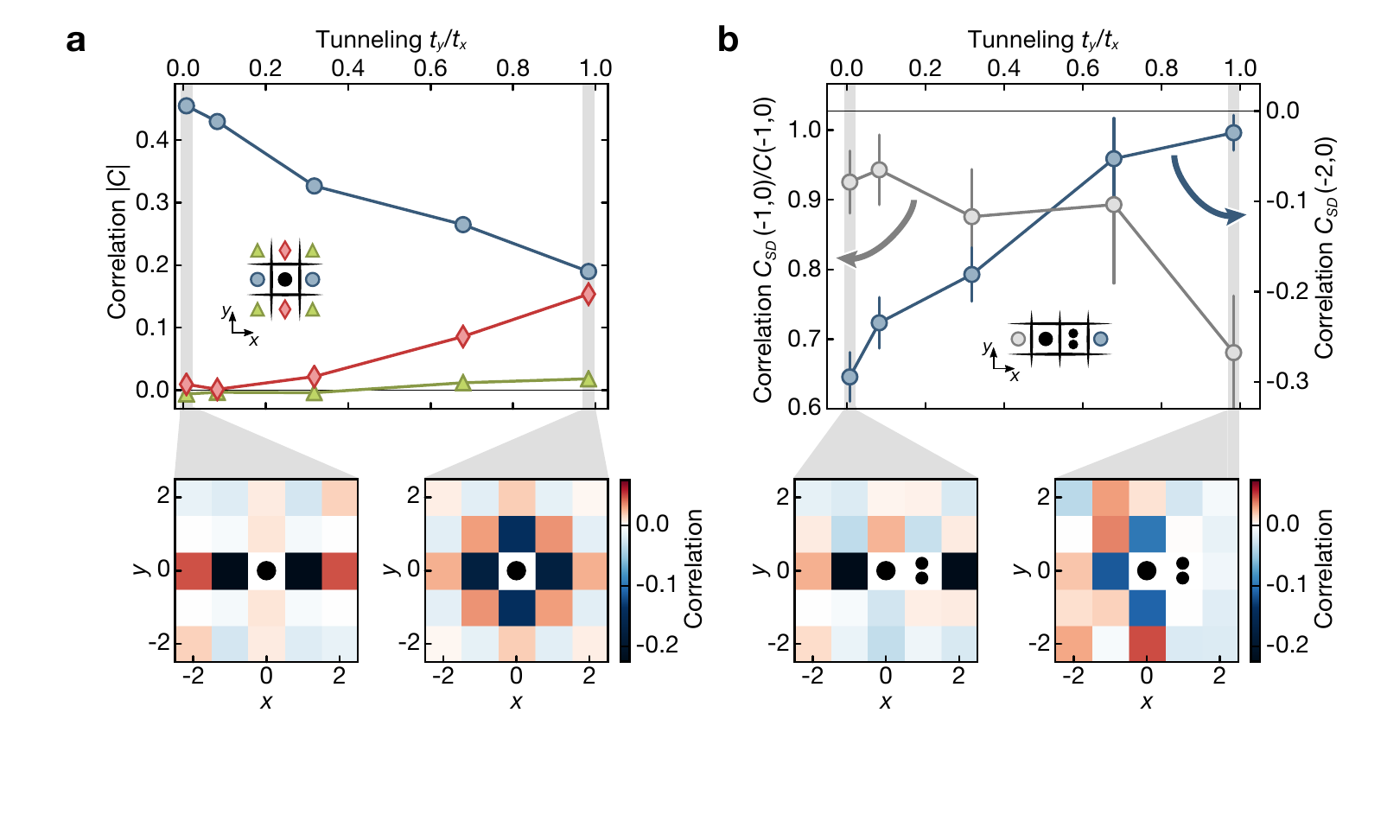}
\caption{\textbf{Spin correlations in the 1D-2D crossover.} \textbf{a}, Spin
correlations $C(x,y)$ as a function of the ratio $t_y/t_x$: $|C(1,0)|$ (blue
circles), $|C(0,1)|$ (red diamonds) and $C(1,1)$ (green triangles) at
$U/t_x$=14. The spin correlations along $x$ decrease as spin correlations
develop in $y$-direction. The pictures below show the 2D spin correlations
amplitudes $C(x,y)$ in the 1D (left) and 2D (right) limits. \textbf{b}, Spin
correlations across doublons $C_{SD}(2,0)$ (blue) and next to it $C_{SD}(-1,0)/C(-1,0)$ (gray) along the $x$ direction.
Antiferromagnetic correlations across doublons, at the origin of
incommensurate correlations in the 1D limit, are strongly suppressed as spin
correlations develop in 2D. In this limit the delocalization of doublons also leads to a reduction of antiferromagnetic correlations on neighbouring sites, indicating the formation of a magnetic polaron. Figures below show the spin correlations $C_{SD}(x,y)$ between
sites $(0,0)$ and $(x,y)$ conditioned on finding a doublon
 on site $(1,0)$ in the 1D (left) and 2D (right) case.}
\label{fig:crossover}
\end{figure*}

Incommensurate spin correlations are also expected to appear in the
one-dimensional Hubbard model when a spin-imbalance is introduced.
To isolate the effect of polarization
from the influence of doping, we consider the two-point spin correlations $C(\tilde{x})=4\langle S^z_{\tilde{i}}
S^z_{\tilde{i}+\tilde{x}} \rangle$ in squeezed space obtained by removing holes and doublons from the
chain in post-analysis~\cite{kruis2004}. In squeezed space~\cite{ogata1990,
woynarovich1982} and for large $U/t_x$, the system is described by a spin-1/2
antiferromagnetic Heisenberg model at a polarization $m=S^z/N_s$, where $N_s$
is the number of singly occupied sites (see Supplementary Information). For the
Heisenberg chain, Luttinger liquid theory predicts at large distances
incommensurate spin correlations linear in the polarization~\cite{giamarchi2004} $m$: \begin{equation} C(\tilde{x}) \simeq A_{m}
e^{-\frac{\tilde{x}}{\xi_{m}}} \cos({\pi (1-2m) \tilde{x}}) \end{equation}
where $A_m,\, \xi_m$ are the magnetization and temperature-dependent amplitude
and correlation length. The SDW wavelength measured by $C(\tilde{x})$ is thus
expected to increase away from $m=0$. In Fig.~\ref{fig:sdw_vs_m}a we show
$C(\tilde{x})$ for two polarizations of the chain $m=-0.12$ and 0. We observe
first a strong decrease of the amplitude of the spin correlations at fixed
distance for finite $m$ compared to $m=0$. This behavior is expected from
conformal field theory which predicts the exponential decay to be stronger due
to a larger zero-temperature critical exponent~\cite{bogoliubov1986}. To reveal
the wave vector of the SDW we computed the Fourier transform of the
spin correlations in squeezed space
$C(\tilde{k})=\mathcal{F}\{C(\tilde{x})\}$ which
qualitatively shows a linear decrease of the wave vector with $m$ (Fig.~\ref{fig:sdw_vs_m}b). We also compare the squeezed space spin correlations at
fixed distance to QMC calculations at half filling for $T/t_x=0.25$ of the
Hubbard chain (Fig.~\ref{fig:sdw_vs_m}c). The good agreement between experiment
and theory validates the use of the squeezed space concept away from $m=0$. 
We attribute the remaining small discrepancy at short distances to our
finite detection efficiency ($97\,\%$), which leads to wrongly detected holes
resulting in an error in the construction of squeezed space. We also expect a
small bias towards lower correlations due to the analysis in squeezed space,
where contributions from lower density areas show smaller
correlations~\cite{hilker2017}.
Similar to the doped case~\cite{hilker2017}, we now study the microscopic origin of these
incommensurate spin correlations. We analyze the spin environment around the
majority spins $C_{e}(\tilde{x})=4\langle S^z_{\tilde{i}} S^z_{\tilde{i}+\tilde{x}}
\rangle_{S^z \sigma_{\tilde{i}+1}>0}$ 
by measuring the conditional expectation value of the spin correlations in
squeezed space for distances $\tilde{x}\geq2$. The correlations are conditioned on the spin $\sigma$
on site $\tilde{i}+1$ being parallel to the chain magnetization $S^z$ (Fig.~\ref{fig:sdw_vs_m}a). We
observe that the sign of the oscillating part in the spin correlations across majority spins changes
compared to the unpolarized case revealing that excess spins act as delocalized domain
walls for the antiferromagnetic order (Fig.~\ref{fig:sdw_vs_m}a). Thus, their
main effect is to increase the distance between antiferromagnetically
correlated spins resulting in an increase of the SDW wavelength as measured by
$C(\tilde{x})$. To formally connect this phenomenon to the effect of doping one
can consider the excess spins $N_\uparrow-N_\downarrow=N_e$ and write the
polarization as $m=n_e/2$, where $n_e$ is the excess spin density. This leads
to a variation of the SDW with distance proportional to $\cos(\pi \bar{n}_e
x)$, where $\bar{n}_e=1-n_e$, in direct analogy to Eq.~\ref{density_eq}
underlining the similarity between the effects of doping and polarization. 
Polarized synthetic Hubbard models have recently also been studied in two
dimensions and the emergence of anisotropic spin correlations has been
observed~\cite{brown2017}.


We now explore the evolution of the spin correlations in the 1D-2D
crossover, a situation relevant to quasi-1D antiferromagnets~\cite{lake2005}.
Whereas in 1D there is no magnetic energy cost associated with the
delocalization of holes and doublons, this phenomenon is expected to breakdown
in higher dimensions. In a 2D antiferromagnetic background the motion of holes
and doublons leads to strings of flipped spins resulting in the confinement of
spin and charge~\cite{brinkman1970,dagotto1994}. The spin correlations around
doublons and holes are thus expected to show qualitative differences in the
crossover from 1D to 2D~\cite{dagotto1994}. We prepared 2D clouds with up to 70
atoms and studied spin correlations while varying $t_y/t_x$ between 0 and 1
keeping $U/t_x=14$ constant (see Supplementary Information). When increasing
$t_y/t_x$, we first observe a decrease in the amplitude of the spin
correlations $C(x,y)=4 \langle S^z_{i,j} S^z_{i+x,j+y}  \rangle_{\scalebox{0.65}{\newmoon}_{i,j}\scalebox{0.65}{\newmoon}_{i+x,j+y}}$ along $x$ and the emergence of spin correlations in the transverse
directions (Fig.~\ref{fig:crossover}a)~\cite{greif2015}. This decrease is
expected even at zero temperature and half filling, where the nearest-neighbor
spin correlations $C(1)$ change from $-0.6$ to $-0.36$ due to the higher
coordination number modifying the quantum fluctuations~\cite{parsons2016}.

Next we study the magnetic environment around doublons in the dimensional crossover from 1D to 2D through $C_{SD}(x,y)=4 \langle S^z_{i,j} S^z_{i+x,j+y}  \rangle_{\scalebox{0.65}{\newmoon}_{i,j}\scalebox{0.65}{\fullmoon}_{i+1,j}\scalebox{0.65}{\newmoon}_{i+x,j+y}}$ 
where the empty circle denotes a doublon located at site $(i+1,j)$ (see Supplementary Information).
We find that the spin correlations across doublons, $C_{SD}(2,0)$, are strongly suppressed while 2D spin correlations develop, which is in stark contrast to the 1D case (Fig.~\ref{fig:crossover}b). Due to the
harmonic confinement the few double occupancies are located in the center of
the trap, where the average density is highest and where magnetic correlations
are expected to compete with doublon delocalization. In addition to the vanishingly
small antiferromagnetic correlations across doublons, we observe a reduction of the
nearest-neighbor spin correlations in its vicinity $C_{SD}(-1,0)/C(-1,0)$ to about $70\%$ compared to
the undoped case (Fig.~\ref{fig:crossover}b). This indicates the formation of a
magnetic polaron~\cite{dagotto1994}, which in the extreme limit $U/t\to \infty$
corresponds to the Nagaoka polaron~\cite{white2001}.


Through the direct simultaneous measurement of both density and spin in the
doped and spin-imbalanced 1D Hubbard model, we shed light onto the connection
between incommensurate spin correlations and the microscopic degrees of
freedom. The spin environment around doublons was found to differ drastically
in the 1D and 2D cases calling for further experimental studies of the
formation of magnetic polarons in homogeneous
systems~\cite{dagotto1994,grusdt2017}. Another interesting extension of this
work is the study of spin correlations as a function of the number of coupled
chains where the parity of the latter is predicted to lead to striking
differences between even and odd cases similar to the problem of half-integer
and integer spin chains~\cite{dagotto1996,white1997}. At low enough temperature
the study of spin and density correlations in hole-doped coupled chains is also
expected to reveal a binding of holes to form stripes which directly extends
the domain-wall concept discussed here to 2D~\cite{white1998}. A study of such
effects through quantum gas microcopy can offer new microscopic insights into
the physics of the doped repulsive Hubbard model.

\section*{Supplementary Information:}

\setcounter{figure}{0}
\renewcommand\thefigure{S\arabic{figure}}    

\section*{Ultracold lattice gas preparation}
The experimental protocol used in the experiments reported here closely
followed our previous work~\cite{hilker2017}. Our experiments started with a
degenerate spin mixture of $^6$Li atoms in the lowest two Zeeman states
$\ket{\pm}=\ket{F=1/2,m_F=\pm1/2}$ trapped in a single plane of a vertical
optical lattice. The lattice spacing was $3.1\,\mu \mathrm{m}$ and the depth
$17\,E^z_r$ (resp. $27\,E^z_r$) in the 1D (crossover) case, where
$E^i_r=h^2/8md_i^2$ is the recoil energy, $m$ the atomic mass and $d_i$ the
lattice spacing along direction $i$. The total atom number $N$ of the cloud was
tuned by varying the depth of a radial trap at the endpoint of the evaporative
cooling procedure~\cite{omran2015}. To simulate the single-band one-dimensional
(1D) Hubbard model, we first prepared 1D systems by ramping up the large
spacing component ($d_y=2.3\,\mu\mathrm{m}$) of an optical superlattice in $y$-direction. The lattice was ramped linearly in two steps, first, to $15\,E^y_r$
in $55\,\mathrm{ms}$ and then to $27\,E^y_r$ in $45\,\mathrm{ms}$, which
results in a final transverse tunneling of $t_y=h\times1.3\,\mathrm{Hz}$. With
a delay of $10\,\mathrm{ms}$ with respect to the start of the $y$-lattice ramp, the lattice
along the tubes ($x$-direction, spacing $d_x=1.15\mu m$)  was turned on. The
chosen ramp was again composed of two linear parts, the first was a ramp to
$3\,E^x_r$ in $45\,\mathrm{ms}$, the second to $5\,E^x_r$ in $55\,\mathrm{ms}$.
The scattering length was simultaneously increased from $530\,a_B$ to
$2000\,a_B$ using a magnetic offset field close to the Feshbach resonance
located at $834\,\mathrm{G}$. At the end of the ramps the tunneling along the
Hubbard chains reached $t_x=h\times 400\,\mathrm{Hz}$ and the onsite
interaction $U=h\times2.9\,\mathrm{kHz}$. The latter is calculated from the
ground band Wannier functions neglecting higher band
corrections~\cite{buchler2010}. The corresponding final superexchange coupling
was $J_x=4t_x^2/U=h\times220\,\mathrm{Hz}$.

To explore the Hubbard model in the crossover from 1D to 2D we first ramped up
the large spacing component of the superlattice in $y$-direction to
$0.2\,E^y_r$ in $60\,\mathrm{ms}$ and then to depths varying between $5\,E^y_r$
and $\,27E^r$ in $220\,\mathrm{ms}$. The $x$-lattice ramp to depths varying between $9E^x_r$ and $10.6E^x_r$ in
$280\,\mathrm{ms}$ started simultaneously with the second part of the $y$-lattice ramp.
The magnetic offset field was adjusted to maintain a constant ratio $U/t_x=14$
at the end of the ramps. A local Stern-Gerlach detection
technique~\cite{boll2016} operating at a transverse magnetic field gradient of
$95\,\mathrm{G/cm}$ was used to detect both the spin and the density on each
lattice site with a fidelity of $97\%$.

\begin{figure}
\centering
\includegraphics{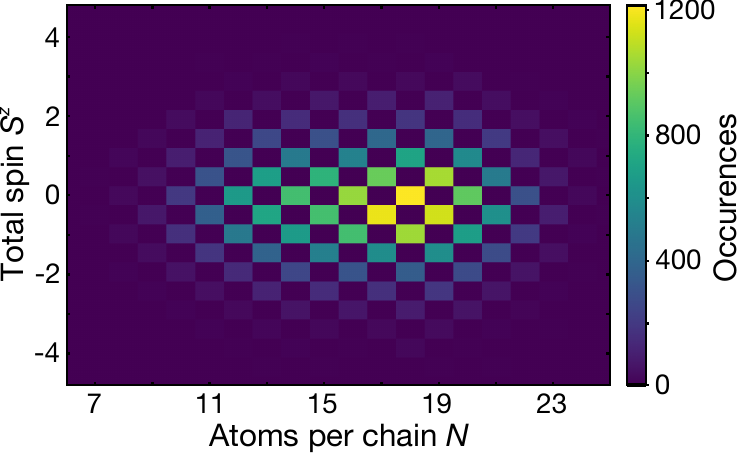}
\caption{\textbf{Chain statistics.} Hubbard chain statistics for a typical
dataset containing $5240$ shots. The total spin $S^z$ and total atom number $N$
of individual Hubbard chains are conserved quantities of the Hamiltonian for
each experimental run. However, they fluctuate for different experimental
realizations allowing to explore individually the effects of doping and
polarization through data grouping.}
\label{fig:histogram}
\end{figure}

\begin{figure*}
\centering
\includegraphics{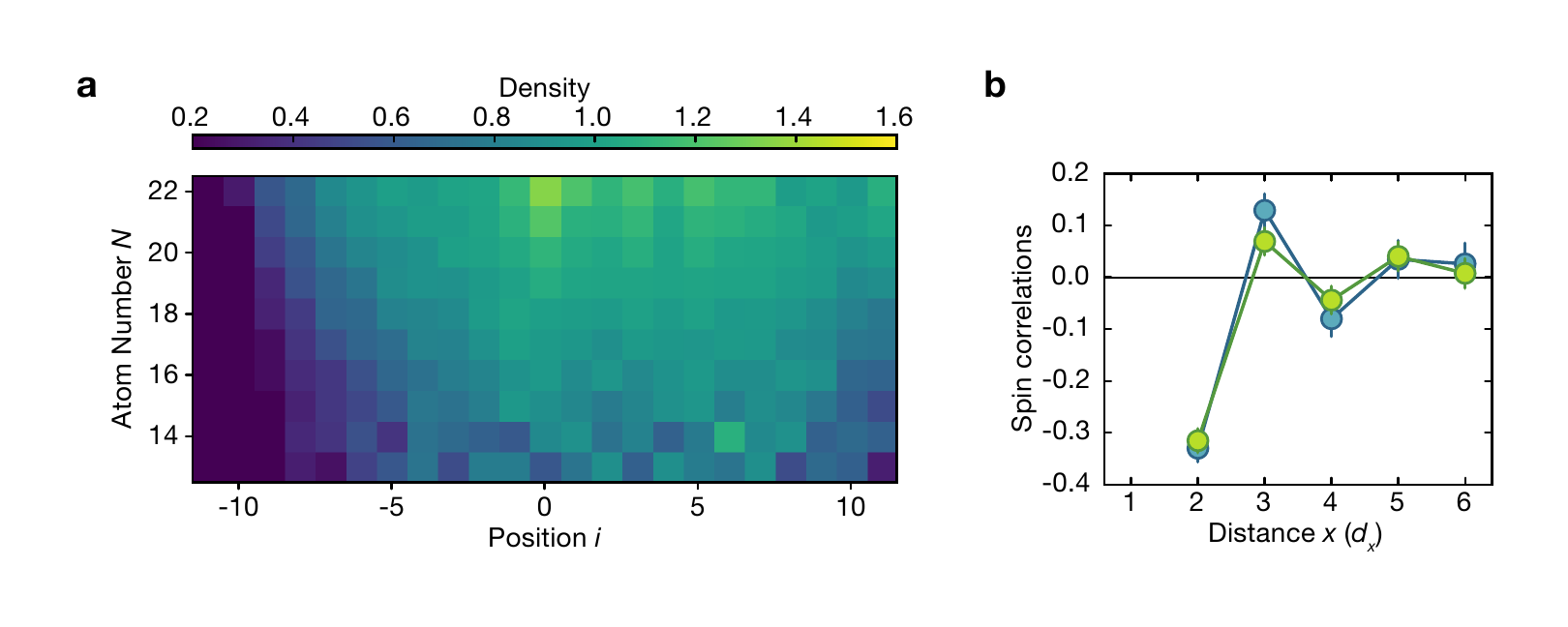}
\caption{\textbf{Density properties of the 1D clouds.} \textbf{a}, Density
profiles $n_0(i,N)$ of the chain located at the center in $y$-direction
($j=0$). \textbf{b}, Antiferromagnetic spin correlations across a hole fixed at
$d_x=1$ (green) or a doublon (blue) measured by $C^{dw}(x)$. The correlation
signal is shifted by the hole which is the microscopic origin of the
incommensurate spin correlations away from half filling in the spin-balanced
case.}
\label{fig:density}
\end{figure*}

\section*{Data analysis}

Thanks to our local access to both spin and occupation on each lattice site in
a single experimental run, we can group each Hubbard chain data by
$\{j,N,S^z\}$, where $j$ is the coordinate of the Hubbard chain in $y$-direction, $N$ and $S^z$ are total atom number and spin. This allows us to
explore different filling and spin sectors (Fig.~\ref{fig:histogram}).
To study the effect of doping on spin correlations we only analyzed the data in
the $S^z=0$ sector. The density profile along $x$ is inhomogeneous and
dependent on $N$ and $j$, which is caused by the underlying harmonic
confinement  of $\omega=2\pi\times 200(20)$Hz. For each pair $\{j,N\}$ we
computed a mean density profile $n_j(i,N)$ by averaging the occupation on each
site $i$ over different experimental realizations (Fig.~\ref{fig:density}). The
reported density at which spin correlations between sites $i$ and $i+x$ have
been analyzed is the mean density between the two points
$n_j(i,x,N)=\frac{1}{x}\sum_{k=i}^{i+x}n_j(k,N)$.

To highlight the oscillatory behavior of the spin correlations as a function of
density we considered the two-point spin correlations between sites $i$ and
$i+x$, conditioned on having single occupancies on these sites for each pair
$\{j,N\}$: 
\begin{equation} 
C_{i,j,N}(x) = 
4\langle S^z_iS^z_{i+x}\rangle_{{\scalebox{0.65}{\newmoon}_i,\scalebox{0.65}{\newmoon}_{i+x},j,N}}
+c(N)
\end{equation} 
where
the filled circles denote single occupancy, the brackets an average over
experimental runs and $c(N)$ is a finite size offset that depends on the atom
number $N$ and temperature, which we experimentally found to be well described
by $c(N)=\frac{1}{N-1}-0.04(5)$~\cite{hilker2017}. We also analyzed the data in
terms of the corresponding connected correlator and found them to agree with
the non-connected version above within statistical uncertainty. This check was
performed also for all other non-connected correlators we use in this
manuscript.
Due to the absence of
density-density correlations beyond $d_x=1$, this correlation function can be
understood as being a renormalized 2-points spin correlation
$C_{i,j,N}(x)\simeq  \frac{4\langle S^z_iS^z_{i+x}\rangle_{j,N}
-c(N)}{n_j(i,x,N)^2}$~\cite{hilker2017}. We finally grouped all
the $C_{i,j,N}(x)$ by their density $n_j(i,x,N)$ in bins of width $\Delta n =
0.1$ to compute the average spin correlation $C(x)$ for each $n$ shown in
Fig.~2 of the main text.\\

The microscopic origin of the incommensurate SDW is revealed by the spin
correlations across holes and double occupancies shown in
Fig.~\ref{fig:density}b: 
\begin{equation} 
C^{dw}_{i,y,N}(x) = 
4\langle S^z_iS^z_{i+x}\rangle_{{\scalebox{0.65}{\newmoon}_i,\scalebox{0.65}{\fullmoon}_{i+1}\scalebox{0.65}{\newmoon}_{i+x},y,N}}
+ c(N),
\end{equation} 
where filled circles denote the condition of having single occupancies on sites
$i,i+x$ and the empty circles denote a doublon or a hole on site $i+1$. The
brackets indicate averaging over all experimental realizations, in which these
conditions are fulfilled. Both the holes and the doublons displace the spin
correlations leading to an increase of their wavelength.

\begin{figure}[t]
\centering
\includegraphics[width=\columnwidth]{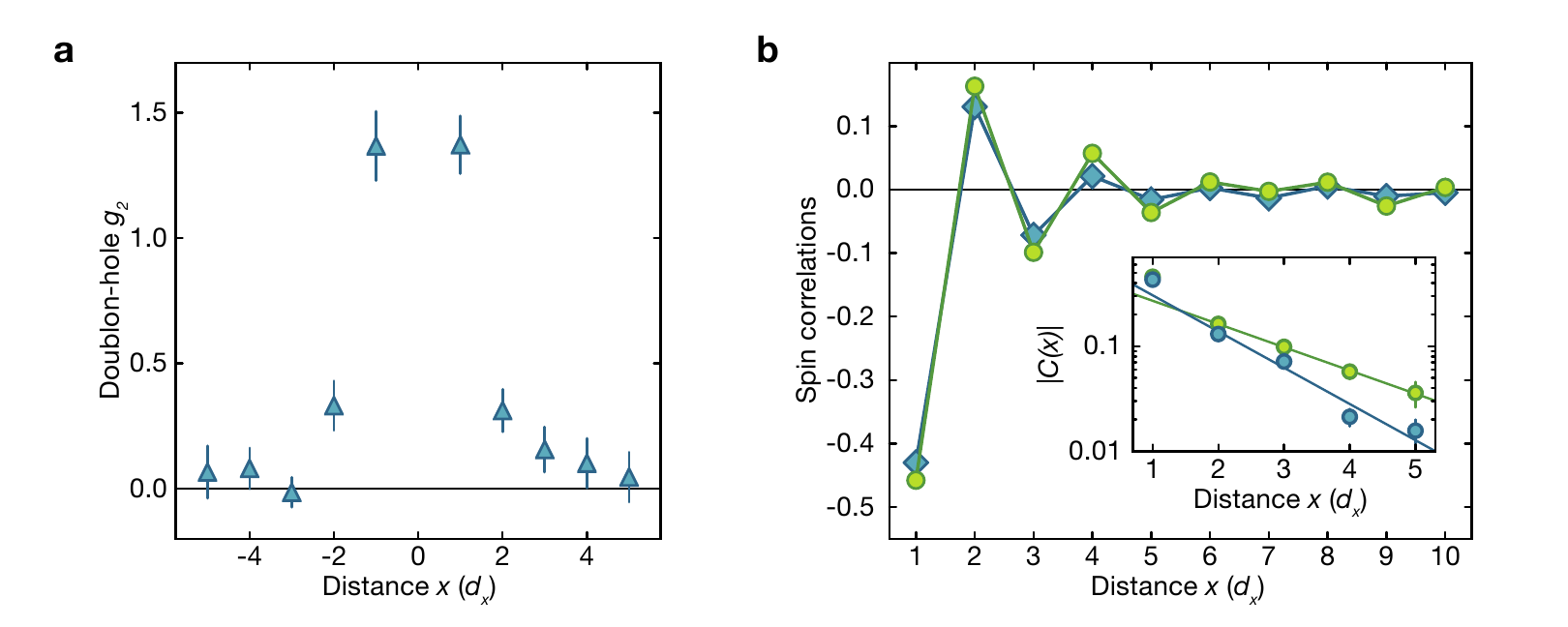}
\caption{\textbf{Spin correlations in squeezed space.} \textbf{a},
Doublon-Hole correlations measured by $g_2(x)$. The strong bunching at $|x|=1$
reveals neighboring doublon-hole pairs as mostly stemming from quantum
fluctuations. This justifies our extension of the squeezed space concept away
from $U\to \infty$.
\textbf{b}, Spin correlations in the zero magnetization sector in the center of
the cloud. The averaging over different polarization (blue) results in a faster
decay of the spin correlations with distance $x$ in squeezed space compared to
the $S^z=0$ sector (green). Exponential fits of the correlation envelope for
distances $x=2,\dots,6$ yield $\xi_{avg}=1.3(1)$ without magnetization
post-selection and $\xi_{0}=2(1)$ in the $S^z=0$ sector.}
\label{fig:magnetization}
\end{figure}

To separate the effect of polarization on the spin correlations from the charge
sector, we studied spin correlations in squeezed space. Here, we extend the
squeezed space concept to finite $U$ by removing doublons and holes only when
these are not nearest-neighbors. The latter condition is supported by the
strong doublon-hole bunching at $d=1$ measured by $g_2(x)=-1+\langle
d_0h_x\rangle/ \langle d_0\rangle\langle h_x\rangle$ (see
Fig.~\ref{fig:magnetization}a), which we attribute to quantum fluctuations. The
full dataset in this measurement consisted of $6474$ experimental runs, in
which we prepared the chains close to half filling in the center. This lead to
chains with up to $N=15$ atoms. We decided to use the squeezed space analysis
instead of post-selecting the data to the zero hole and doublon sector to improve our
statistics. Within statistical uncertainties, the post-selected data is
consistent with these squeezed state results.

The spin correlations in squeezed space, indicated by $\tilde{i}$ and $\tilde{x}$, are defined as:
\begin{equation}
C_{\tilde{i},j,N_s,S^z}(\tilde{x}) =
4\langle S^z_{\tilde{i}}S^z_{\tilde{i}+\tilde{x}}\rangle
+ c_{sq}(N_s, S^z),
\end{equation} 
where $N_s$ is the number of singly occupied sites including nearest neighbor
doublon-hole pairs. We again take into account a finite size offset
$c_{sq}(N_s,S^z)=\frac{1}{N_s-1}-\frac{4(S^z)^2}{N_s(N_s-1)}-0.05(5)$ in
analogy to $c(N)$~\cite{hilker2017}.
The magnetization of the effective Heisenberg chain in squeezed space is
defined as  $m=S^z/N_s$ . We grouped all the
$C_{\tilde{i},j,N_s,S^z}(\tilde{x})$ by their polarization $m$ in bins of width
$\Delta m = 0.04$ to compute the average spin correlation $C$ shown in Fig.~3
of the main text. 
The Fourier transform shown in Fig.~3b of the main text has been calculated
without the nearest-neighbor correlations (i.e. for $\tilde{x}\geq2$) to avoid
short distance effects.\\

Similar to the hole and doublon case, the microscopic origin of the
polarization dependence of the SDW wavelength can be revealed by the oscillating part in the spin
correlations across majority spins:
\begin{equation} 
C^{e}_{\tilde{i},j,S^z,N_s}(\tilde{x}) = 
4\langle S^z_{\tilde{i}} S^z_{\tilde{i}+\tilde{x}} \rangle_{S^z \sigma_{\tilde{i}+1}>0}
-c_{sq,2}(N_s,S^z), 
\end{equation} 
where the spin $\sigma$ on site $\tilde{i}+1$ is parallel to the total magnetization in the chain $S^z$ and $c_{sq,2}(N_s,S^z)=\frac{2}{(N_s-1)(N_s-2)}\left(2(S^z)^2-2|S^z|-N_s/2+1\right)$ is the corresponding finite size and finite magnetization offset for uncorrelated spins.  The brackets indicate the average over all experimental realization
where this condition is fulfilled. We finally averaged the correlator over
$\{\tilde{i},j,S^z,N_s\}$ and binned $m=0.12\pm0.02$ to obtain
$C_{e}(\tilde{x})$ presented on Fig.~3 of the main text. The sign change of the
spin correlations at $\tilde{x} \geq 2$ reveal that the excess magnetization is
carried by delocalized spinons. These stretch spin correlations and lead to
incommensurate magnetism when $m\neq 0$.

\begin{figure*}
\centering
\includegraphics{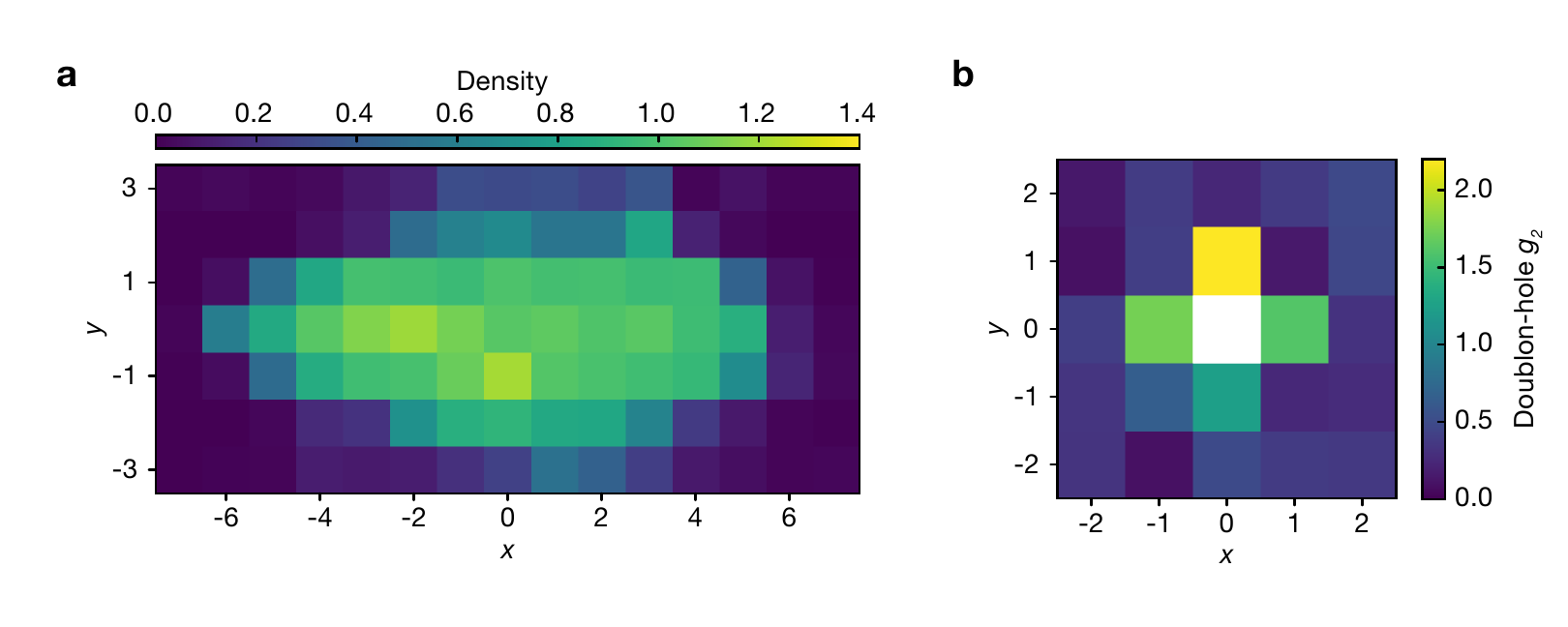}
\caption{\textbf{Properties of the prepared 2D clouds.}  \textbf{a}, Density
distribution for $t_y/t_x=1$. \textbf{b}, Doublon-hole correlations
$g_2(\vec{r})$. The	strong bunching of the doubon-hole correlations $g_2(\vec{r})$ at $|\vec{r}|=1$ justifies to discard
outcomes where holes and doublons are found nearby when studying the effects of
doping.}
\label{fig:crossover_sup}
\end{figure*}

In the dimensional crossover and 2D regime we prepared anisotropic samples
consisting of about five coupled Hubbard chains (Fig.~\ref{fig:crossover_sup}).
Similarly to the 1D case, the spin correlations were calculated on singly
occupied sites through $C_{\vec{n}}({\vec{r}})=4 \langle
S^z_{\vec{n}}S^z_{\vec{n}+\vec{r}}\rangle_{{\scalebox{0.65}{\newmoon}_{\vec{n}},\scalebox{0.65}{\newmoon}_{\vec{n}+\vec{r}}}}$ averaged over all sites $\vec{n}$. When studying spin correlations around double occupancies and holes in the
crossover, we took care to minimize biasing of the correlator by a possibly
distorted magnetic background around quantum fluctuation induced doublon-hole
pairs. The strong bunching observed in $g_2(\vec{r})$ (see
Fig.~\ref{fig:crossover_sup}) on nearest neighbor scale identifies a strong
contribution of quantum fluctuations to these. Hence, we discarded from the analysis of the spin correlations any doublons having one of its nearest neighbor unoccupied.
\section*{Quantum Monte-Carlo calculations}

The Quantum Monte Carlo (QMC) results shown in the present work are obtained in
a similar fashion as those found in \cite{boll2016}.  Simulating the
fermionic system without  sign problem is made possible by the mapping between
the one-dimensional fermionic Hubbard model and a system of two hard-core
bosonic species with on-site interspecies interactions~\cite{jordan1928}.

We make use of the worm algorithm~\cite{prokofev1998} in the implementation of
Ref.~\cite{pollet2007}. This algorithm exhibits a linear scaling in the system
volume when simulating the resulting bosonic model.  The spin $S_i$ at site $i$
of the fermionic model maps onto a diagonal observable with respect to the Fock
basis $ \{ \vert \ldots, n_j ,\ldots \rangle \} $ of the bosonic model,
proportional to the differences in the occupation numbers of the bosonic
particles at the same site.

The simulations were all carried out in the grand-canonical ensemble. The
system consisted of a homogeneous lattice of $L=20$ sites with hard-wall
boundary conditions. This size was checked to be already large enough to avoid
finite size corrections. Note however that the correlations are affected by an
unavoidable systematic offset which scales as $1/N$ and which was corrected for
in the analysis as explained above.

To better mimic the measurement procedure of the actual experiment, we saved
the raw QMC configurations and performed the analysis off-line. In this
process, care must be taken to make sure that subsequent configurations are
decorrelated. A further blocking and jackknife estimation was used to rule out
any residual correlation.

The off-line analyses were subject to the same filtering procedures for the
occupation and magnetization sector as it was done in the experimental
procedure. In order to gather enough statistics for the different values of the
density accessible in the experiment, we tuned the chemical potentials of the
two bosonic species so as to have symmetric mixtures with total density $n$
between $0.4$ and $1.2$.  Similarly, in order to more efficiently collect
statistics in nonzero polarization sectors, we tuned the chemical potentials in
an anti-symmetric way with respect to the symmetric half-filling condition,
i.e., $\mu_{1,2} = \mu_{\rm hf} \pm \Delta\mu$, with $\mu_{\rm hf}$ the
chemical potential of the half filled symmetric binary mixture and $\Delta \mu$
chosen so as to access chains with polarizations $m$ between $-0.2$ and $0.2$.

\bigskip

\begin{acknowledgments}
\textbf{Acknowledgments:} We thank T. Giamarchi for illuminating exchanges on
incommensurate magnetism, D. Huse, A. Recati, E. Demler and F. Grusdt for helpful
discussions and P. Sompet for critical reading of the manuscript. Financial
support was provided by the Max Planck Society (MPG), the European Union
(UQUAM, QSIMGAS, MIR-BOSE) and J.K. acknowledges funding from the Hector Fellow
Academy.
\end{acknowledgments}

\bigskip

\textbf{Author Contributions:} 
G.~S., T.~A.~H., J.~K., J.~V., I.~B. and C.~G. planned the experiment, analyzed
and discussed the data. J.~N. and L.~P. performed the QMC simulations. All
authors contributed to the interpretation of the data and the writing.

\bigskip

\textbf{Materials and Correspondence:} Correspondence and requests for
materials should be addressed to guillaume.salomon@mpq.mpg.de

\bibliography{bibliography}

\begin{thebibliography}{36}%
\makeatletter
\providecommand \@ifxundefined [1]{%
 \@ifx{#1\undefined}
}%
\providecommand \@ifnum [1]{%
 \ifnum #1\expandafter \@firstoftwo
 \else \expandafter \@secondoftwo
 \fi
}%
\providecommand \@ifx [1]{%
 \ifx #1\expandafter \@firstoftwo
 \else \expandafter \@secondoftwo
 \fi
}%
\providecommand \natexlab [1]{#1}%
\providecommand \enquote  [1]{``#1''}%
\providecommand \bibnamefont  [1]{#1}%
\providecommand \bibfnamefont [1]{#1}%
\providecommand \citenamefont [1]{#1}%
\providecommand \href@noop [0]{\@secondoftwo}%
\providecommand \href [0]{\begingroup \@sanitize@url \@href}%
\providecommand \@href[1]{\@@startlink{#1}\@@href}%
\providecommand \@@href[1]{\endgroup#1\@@endlink}%
\providecommand \@sanitize@url [0]{\catcode `\\12\catcode `\$12\catcode
  `\&12\catcode `\#12\catcode `\^12\catcode `\_12\catcode `\%12\relax}%
\providecommand \@@startlink[1]{}%
\providecommand \@@endlink[0]{}%
\providecommand \url  [0]{\begingroup\@sanitize@url \@url }%
\providecommand \@url [1]{\endgroup\@href {#1}{\urlprefix }}%
\providecommand \urlprefix  [0]{URL }%
\providecommand \Eprint [0]{\href }%
\providecommand \doibase [0]{http://dx.doi.org/}%
\providecommand \selectlanguage [0]{\@gobble}%
\providecommand \bibinfo  [0]{\@secondoftwo}%
\providecommand \bibfield  [0]{\@secondoftwo}%
\providecommand \translation [1]{[#1]}%
\providecommand \BibitemOpen [0]{}%
\providecommand \bibitemStop [0]{}%
\providecommand \bibitemNoStop [0]{.\EOS\space}%
\providecommand \EOS [0]{\spacefactor3000\relax}%
\providecommand \BibitemShut  [1]{\csname bibitem#1\endcsname}%
\let\auto@bib@innerbib\@empty
\bibitem [{\citenamefont {Giamarchi}(2003)}]{giamarchi2004}%
  \BibitemOpen
  \bibfield  {author} {\bibinfo {author} {\bibfnamefont {T.}~\bibnamefont
  {Giamarchi}},\ }\href@noop {} {\emph {\bibinfo {title} {Quantum {{Physics}}
  in {{One Dimension}}}}}\ (\bibinfo  {publisher} {Clarendon Press},\ \bibinfo
  {year} {2003})\BibitemShut {NoStop}%
\bibitem [{\citenamefont {Greif}\ \emph {et~al.}(2013)\citenamefont {Greif},
  \citenamefont {Uehlinger}, \citenamefont {Jotzu}, \citenamefont {Tarruell},\
  and\ \citenamefont {Esslinger}}]{greif2013}%
  \BibitemOpen
  \bibfield  {author} {\bibinfo {author} {\bibfnamefont {D.}~\bibnamefont
  {Greif}}, \bibinfo {author} {\bibfnamefont {T.}~\bibnamefont {Uehlinger}},
  \bibinfo {author} {\bibfnamefont {G.}~\bibnamefont {Jotzu}}, \bibinfo
  {author} {\bibfnamefont {L.}~\bibnamefont {Tarruell}}, \ and\ \bibinfo
  {author} {\bibfnamefont {T.}~\bibnamefont {Esslinger}},\ }\href {\doibase
  10.1126/science.1236362} {\bibfield  {journal} {\bibinfo  {journal}
  {Science}\ }\textbf {\bibinfo {volume} {340}},\ \bibinfo {pages} {1307}
  (\bibinfo {year} {2013})}\BibitemShut {NoStop}%
\bibitem [{\citenamefont {Hart}\ \emph {et~al.}(2015)\citenamefont {Hart},
  \citenamefont {Duarte}, \citenamefont {Yang}, \citenamefont {Liu},
  \citenamefont {Paiva}, \citenamefont {Khatami}, \citenamefont {Scalettar},
  \citenamefont {Trivedi}, \citenamefont {Huse},\ and\ \citenamefont
  {Hulet}}]{hart2015}%
  \BibitemOpen
  \bibfield  {author} {\bibinfo {author} {\bibfnamefont {R.~A.}\ \bibnamefont
  {Hart}}, \bibinfo {author} {\bibfnamefont {P.~M.}\ \bibnamefont {Duarte}},
  \bibinfo {author} {\bibfnamefont {T.-L.}\ \bibnamefont {Yang}}, \bibinfo
  {author} {\bibfnamefont {X.}~\bibnamefont {Liu}}, \bibinfo {author}
  {\bibfnamefont {T.}~\bibnamefont {Paiva}}, \bibinfo {author} {\bibfnamefont
  {E.}~\bibnamefont {Khatami}}, \bibinfo {author} {\bibfnamefont {R.~T.}\
  \bibnamefont {Scalettar}}, \bibinfo {author} {\bibfnamefont {N.}~\bibnamefont
  {Trivedi}}, \bibinfo {author} {\bibfnamefont {D.~A.}\ \bibnamefont {Huse}}, \
  and\ \bibinfo {author} {\bibfnamefont {R.~G.}\ \bibnamefont {Hulet}},\
  }\href@noop {} {\bibfield  {journal} {\bibinfo  {journal} {Nature}\ }\textbf
  {\bibinfo {volume} {519}},\ \bibinfo {pages} {211} (\bibinfo {year}
  {2015})}\BibitemShut {NoStop}%
\bibitem [{\citenamefont {Parsons}\ \emph {et~al.}(2016)\citenamefont
  {Parsons}, \citenamefont {Mazurenko}, \citenamefont {Chiu}, \citenamefont
  {Ji}, \citenamefont {Greif},\ and\ \citenamefont {Greiner}}]{parsons2016}%
  \BibitemOpen
  \bibfield  {author} {\bibinfo {author} {\bibfnamefont {M.~F.}\ \bibnamefont
  {Parsons}}, \bibinfo {author} {\bibfnamefont {A.}~\bibnamefont {Mazurenko}},
  \bibinfo {author} {\bibfnamefont {C.~S.}\ \bibnamefont {Chiu}}, \bibinfo
  {author} {\bibfnamefont {G.}~\bibnamefont {Ji}}, \bibinfo {author}
  {\bibfnamefont {D.}~\bibnamefont {Greif}}, \ and\ \bibinfo {author}
  {\bibfnamefont {M.}~\bibnamefont {Greiner}},\ }\href {\doibase
  10.1126/science.aag1430} {\bibfield  {journal} {\bibinfo  {journal}
  {Science}\ }\textbf {\bibinfo {volume} {353}},\ \bibinfo {pages} {1253}
  (\bibinfo {year} {2016})}\BibitemShut {NoStop}%
\bibitem [{\citenamefont {Boll}\ \emph {et~al.}(2016)\citenamefont {Boll},
  \citenamefont {Hilker}, \citenamefont {Salomon}, \citenamefont {Omran},
  \citenamefont {Nespolo}, \citenamefont {Pollet}, \citenamefont {Bloch},\ and\
  \citenamefont {Gross}}]{boll2016}%
  \BibitemOpen
  \bibfield  {author} {\bibinfo {author} {\bibfnamefont {M.}~\bibnamefont
  {Boll}}, \bibinfo {author} {\bibfnamefont {T.~A.}\ \bibnamefont {Hilker}},
  \bibinfo {author} {\bibfnamefont {G.}~\bibnamefont {Salomon}}, \bibinfo
  {author} {\bibfnamefont {A.}~\bibnamefont {Omran}}, \bibinfo {author}
  {\bibfnamefont {J.}~\bibnamefont {Nespolo}}, \bibinfo {author} {\bibfnamefont
  {L.}~\bibnamefont {Pollet}}, \bibinfo {author} {\bibfnamefont
  {I.}~\bibnamefont {Bloch}}, \ and\ \bibinfo {author} {\bibfnamefont
  {C.}~\bibnamefont {Gross}},\ }\href {\doibase 10.1126/science.aag1635}
  {\bibfield  {journal} {\bibinfo  {journal} {Science}\ }\textbf {\bibinfo
  {volume} {353}},\ \bibinfo {pages} {1257} (\bibinfo {year}
  {2016})}\BibitemShut {NoStop}%
\bibitem [{\citenamefont {Cheuk}\ \emph {et~al.}(2016)\citenamefont {Cheuk},
  \citenamefont {Nichols}, \citenamefont {Lawrence}, \citenamefont {Okan},
  \citenamefont {Zhang}, \citenamefont {Khatami}, \citenamefont {Trivedi},
  \citenamefont {Paiva}, \citenamefont {Rigol},\ and\ \citenamefont
  {Zwierlein}}]{cheuk2016}%
  \BibitemOpen
  \bibfield  {author} {\bibinfo {author} {\bibfnamefont {L.~W.}\ \bibnamefont
  {Cheuk}}, \bibinfo {author} {\bibfnamefont {M.~A.}\ \bibnamefont {Nichols}},
  \bibinfo {author} {\bibfnamefont {K.~R.}\ \bibnamefont {Lawrence}}, \bibinfo
  {author} {\bibfnamefont {M.}~\bibnamefont {Okan}}, \bibinfo {author}
  {\bibfnamefont {H.}~\bibnamefont {Zhang}}, \bibinfo {author} {\bibfnamefont
  {E.}~\bibnamefont {Khatami}}, \bibinfo {author} {\bibfnamefont
  {N.}~\bibnamefont {Trivedi}}, \bibinfo {author} {\bibfnamefont
  {T.}~\bibnamefont {Paiva}}, \bibinfo {author} {\bibfnamefont
  {M.}~\bibnamefont {Rigol}}, \ and\ \bibinfo {author} {\bibfnamefont {M.~W.}\
  \bibnamefont {Zwierlein}},\ }\href {\doibase 10.1126/science.aag3349}
  {\bibfield  {journal} {\bibinfo  {journal} {Science}\ }\textbf {\bibinfo
  {volume} {353}},\ \bibinfo {pages} {1260} (\bibinfo {year}
  {2016})}\BibitemShut {NoStop}%
\bibitem [{\citenamefont {Drewes}\ \emph {et~al.}(2017)\citenamefont {Drewes},
  \citenamefont {Miller}, \citenamefont {Cocchi}, \citenamefont {Chan},
  \citenamefont {Wurz}, \citenamefont {Gall}, \citenamefont {Pertot},
  \citenamefont {Brennecke},\ and\ \citenamefont {K\"ohl}}]{drewes2017}%
  \BibitemOpen
  \bibfield  {author} {\bibinfo {author} {\bibfnamefont {J.~H.}\ \bibnamefont
  {Drewes}}, \bibinfo {author} {\bibfnamefont {L.~A.}\ \bibnamefont {Miller}},
  \bibinfo {author} {\bibfnamefont {E.}~\bibnamefont {Cocchi}}, \bibinfo
  {author} {\bibfnamefont {C.~F.}\ \bibnamefont {Chan}}, \bibinfo {author}
  {\bibfnamefont {N.}~\bibnamefont {Wurz}}, \bibinfo {author} {\bibfnamefont
  {M.}~\bibnamefont {Gall}}, \bibinfo {author} {\bibfnamefont {D.}~\bibnamefont
  {Pertot}}, \bibinfo {author} {\bibfnamefont {F.}~\bibnamefont {Brennecke}}, \
  and\ \bibinfo {author} {\bibfnamefont {M.}~\bibnamefont {K\"ohl}},\ }\href
  {\doibase 10.1103/PhysRevLett.118.170401} {\bibfield  {journal} {\bibinfo
  {journal} {Phys. Rev. Lett.}\ }\textbf {\bibinfo {volume} {118}},\ \bibinfo
  {pages} {170401} (\bibinfo {year} {2017})}\BibitemShut {NoStop}%
\bibitem [{\citenamefont {Brown}\ \emph {et~al.}(2017)\citenamefont {Brown},
  \citenamefont {Mitra}, \citenamefont {Guardado-Sanchez}, \citenamefont
  {Schau{\ss}}, \citenamefont {Kondov}, \citenamefont {Khatami}, \citenamefont
  {Paiva}, \citenamefont {Trivedi}, \citenamefont {Huse},\ and\ \citenamefont
  {Bakr}}]{brown2017}%
  \BibitemOpen
  \bibfield  {author} {\bibinfo {author} {\bibfnamefont {P.~T.}\ \bibnamefont
  {Brown}}, \bibinfo {author} {\bibfnamefont {D.}~\bibnamefont {Mitra}},
  \bibinfo {author} {\bibfnamefont {E.}~\bibnamefont {Guardado-Sanchez}},
  \bibinfo {author} {\bibfnamefont {P.}~\bibnamefont {Schau{\ss}}}, \bibinfo
  {author} {\bibfnamefont {S.~S.}\ \bibnamefont {Kondov}}, \bibinfo {author}
  {\bibfnamefont {E.}~\bibnamefont {Khatami}}, \bibinfo {author} {\bibfnamefont
  {T.}~\bibnamefont {Paiva}}, \bibinfo {author} {\bibfnamefont
  {N.}~\bibnamefont {Trivedi}}, \bibinfo {author} {\bibfnamefont {D.~A.}\
  \bibnamefont {Huse}}, \ and\ \bibinfo {author} {\bibfnamefont {W.~S.}\
  \bibnamefont {Bakr}},\ }\href {\doibase 10.1126/science.aam7838} {\bibfield
  {journal} {\bibinfo  {journal} {Science}\ }\textbf {\bibinfo {volume}
  {357}},\ \bibinfo {pages} {1385} (\bibinfo {year} {2017})}\BibitemShut
  {NoStop}%
\bibitem [{\citenamefont {Dagotto}(1994)}]{dagotto1994}%
  \BibitemOpen
  \bibfield  {author} {\bibinfo {author} {\bibfnamefont {E.}~\bibnamefont
  {Dagotto}},\ }\href {\doibase 10.1103/RevModPhys.66.763} {\bibfield
  {journal} {\bibinfo  {journal} {Rev. Mod. Phys.}\ }\textbf {\bibinfo {volume}
  {66}},\ \bibinfo {pages} {763} (\bibinfo {year} {1994})}\BibitemShut
  {NoStop}%
\bibitem [{\citenamefont {Haldane}(1981)}]{haldane1981a}%
  \BibitemOpen
  \bibfield  {author} {\bibinfo {author} {\bibfnamefont {F.~D.~M.}\
  \bibnamefont {Haldane}},\ }\href@noop {} {\bibfield  {journal} {\bibinfo
  {journal} {Journal of Physics C: Solid State Physics}\ }\textbf {\bibinfo
  {volume} {14}},\ \bibinfo {pages} {2585} (\bibinfo {year}
  {1981})}\BibitemShut {NoStop}%
\bibitem [{\citenamefont {Wen}(1990)}]{wen90}%
  \BibitemOpen
  \bibfield  {author} {\bibinfo {author} {\bibfnamefont {X.~G.}\ \bibnamefont
  {Wen}},\ }\href {\doibase 10.1103/PhysRevB.41.12838} {\bibfield  {journal}
  {\bibinfo  {journal} {Phys. Rev. B}\ }\textbf {\bibinfo {volume} {41}},\
  \bibinfo {pages} {12838} (\bibinfo {year} {1990})}\BibitemShut {NoStop}%
\bibitem [{\citenamefont {Frahm}\ and\ \citenamefont
  {Korepin}(1991)}]{frahm1991}%
  \BibitemOpen
  \bibfield  {author} {\bibinfo {author} {\bibfnamefont {H.}~\bibnamefont
  {Frahm}}\ and\ \bibinfo {author} {\bibfnamefont {V.~E.}\ \bibnamefont
  {Korepin}},\ }\href {\doibase 10.1103/PhysRevB.43.5653} {\bibfield  {journal}
  {\bibinfo  {journal} {Phys. Rev. B}\ }\textbf {\bibinfo {volume} {43}},\
  \bibinfo {pages} {5653} (\bibinfo {year} {1991})}\BibitemShut {NoStop}%
\bibitem [{\citenamefont {Cardy}(1996)}]{cardy1996}%
  \BibitemOpen
  \bibfield  {author} {\bibinfo {author} {\bibfnamefont {J.}~\bibnamefont
  {Cardy}},\ }\href@noop {} {\emph {\bibinfo {title} {{Scaling and
  Renormalization in Statistical Physics}}}}\ (\bibinfo  {publisher} {Cambridge
  University Press},\ \bibinfo {year} {1996})\BibitemShut {NoStop}%
\bibitem [{\citenamefont {Bockrath}\ \emph {et~al.}(1999)\citenamefont
  {Bockrath}, \citenamefont {Cobden}, \citenamefont {Lu}, \citenamefont
  {Rinzler}, \citenamefont {Smalley}, \citenamefont {Balents},\ and\
  \citenamefont {McEuen}}]{bockrath1999}%
  \BibitemOpen
  \bibfield  {author} {\bibinfo {author} {\bibfnamefont {M.}~\bibnamefont
  {Bockrath}}, \bibinfo {author} {\bibfnamefont {D.~H.}\ \bibnamefont
  {Cobden}}, \bibinfo {author} {\bibfnamefont {J.}~\bibnamefont {Lu}}, \bibinfo
  {author} {\bibfnamefont {A.~G.}\ \bibnamefont {Rinzler}}, \bibinfo {author}
  {\bibfnamefont {R.~E.}\ \bibnamefont {Smalley}}, \bibinfo {author}
  {\bibfnamefont {L.}~\bibnamefont {Balents}}, \ and\ \bibinfo {author}
  {\bibfnamefont {P.~L.}\ \bibnamefont {McEuen}},\ }\href@noop {} {\bibfield
  {journal} {\bibinfo  {journal} {Nature}\ }\textbf {\bibinfo {volume} {397}},\
  \bibinfo {pages} {598} (\bibinfo {year} {1999})}\BibitemShut {NoStop}%
\bibitem [{\citenamefont {Lee}\ \emph {et~al.}(2004)\citenamefont {Lee},
  \citenamefont {Eggert}, \citenamefont {Kim}, \citenamefont {Kahng},
  \citenamefont {Shinohara},\ and\ \citenamefont {Kuk}}]{lee2004}%
  \BibitemOpen
  \bibfield  {author} {\bibinfo {author} {\bibfnamefont {J.}~\bibnamefont
  {Lee}}, \bibinfo {author} {\bibfnamefont {S.}~\bibnamefont {Eggert}},
  \bibinfo {author} {\bibfnamefont {H.}~\bibnamefont {Kim}}, \bibinfo {author}
  {\bibfnamefont {S.-J.}\ \bibnamefont {Kahng}}, \bibinfo {author}
  {\bibfnamefont {H.}~\bibnamefont {Shinohara}}, \ and\ \bibinfo {author}
  {\bibfnamefont {Y.}~\bibnamefont {Kuk}},\ }\href {\doibase
  10.1103/PhysRevLett.93.166403} {\bibfield  {journal} {\bibinfo  {journal}
  {Phys. Rev. Lett.}\ }\textbf {\bibinfo {volume} {93}},\ \bibinfo {pages}
  {166403} (\bibinfo {year} {2004})}\BibitemShut {NoStop}%
\bibitem [{\citenamefont {Stone}\ \emph {et~al.}(2003)\citenamefont {Stone},
  \citenamefont {Reich}, \citenamefont {Broholm}, \citenamefont {Lefmann},
  \citenamefont {Rischel}, \citenamefont {Landee},\ and\ \citenamefont
  {Turnbull}}]{stone2003}%
  \BibitemOpen
  \bibfield  {author} {\bibinfo {author} {\bibfnamefont {M.~B.}\ \bibnamefont
  {Stone}}, \bibinfo {author} {\bibfnamefont {D.~H.}\ \bibnamefont {Reich}},
  \bibinfo {author} {\bibfnamefont {C.}~\bibnamefont {Broholm}}, \bibinfo
  {author} {\bibfnamefont {K.}~\bibnamefont {Lefmann}}, \bibinfo {author}
  {\bibfnamefont {C.}~\bibnamefont {Rischel}}, \bibinfo {author} {\bibfnamefont
  {C.~P.}\ \bibnamefont {Landee}}, \ and\ \bibinfo {author} {\bibfnamefont
  {M.~M.}\ \bibnamefont {Turnbull}},\ }\href {\doibase
  10.1103/PhysRevLett.91.037205} {\bibfield  {journal} {\bibinfo  {journal}
  {Phys. Rev. Lett.}\ }\textbf {\bibinfo {volume} {91}},\ \bibinfo {pages}
  {037205} (\bibinfo {year} {2003})}\BibitemShut {NoStop}%
\bibitem [{\citenamefont {Lake}\ \emph {et~al.}(2005)\citenamefont {Lake},
  \citenamefont {Tennant}, \citenamefont {Frost},\ and\ \citenamefont
  {Nagler}}]{lake2005}%
  \BibitemOpen
  \bibfield  {author} {\bibinfo {author} {\bibfnamefont {B.}~\bibnamefont
  {Lake}}, \bibinfo {author} {\bibfnamefont {D.~A.}\ \bibnamefont {Tennant}},
  \bibinfo {author} {\bibfnamefont {C.~D.}\ \bibnamefont {Frost}}, \ and\
  \bibinfo {author} {\bibfnamefont {S.~E.}\ \bibnamefont {Nagler}},\
  }\href@noop {} {\bibfield  {journal} {\bibinfo  {journal} {Nature Materials}\
  }\textbf {\bibinfo {volume} {4}},\ \bibinfo {pages} {329} (\bibinfo {year}
  {2005})}\BibitemShut {NoStop}%
\bibitem [{\citenamefont {Klanj\ifmmode~\check{s}\else \v{s}\fi{}ek}\ \emph
  {et~al.}(2008)\citenamefont {Klanj\ifmmode~\check{s}\else \v{s}\fi{}ek},
  \citenamefont {Mayaffre}, \citenamefont {Berthier}, \citenamefont
  {Horvati\ifmmode~\acute{c}\else \'{c}\fi{}}, \citenamefont {Chiari},
  \citenamefont {Piovesana}, \citenamefont {Bouillot}, \citenamefont {Kollath},
  \citenamefont {Orignac}, \citenamefont {Citro},\ and\ \citenamefont
  {Giamarchi}}]{klanjsek2008}%
  \BibitemOpen
  \bibfield  {author} {\bibinfo {author} {\bibfnamefont {M.}~\bibnamefont
  {Klanj\ifmmode~\check{s}\else \v{s}\fi{}ek}}, \bibinfo {author}
  {\bibfnamefont {H.}~\bibnamefont {Mayaffre}}, \bibinfo {author}
  {\bibfnamefont {C.}~\bibnamefont {Berthier}}, \bibinfo {author}
  {\bibfnamefont {M.}~\bibnamefont {Horvati\ifmmode~\acute{c}\else
  \'{c}\fi{}}}, \bibinfo {author} {\bibfnamefont {B.}~\bibnamefont {Chiari}},
  \bibinfo {author} {\bibfnamefont {O.}~\bibnamefont {Piovesana}}, \bibinfo
  {author} {\bibfnamefont {P.}~\bibnamefont {Bouillot}}, \bibinfo {author}
  {\bibfnamefont {C.}~\bibnamefont {Kollath}}, \bibinfo {author} {\bibfnamefont
  {E.}~\bibnamefont {Orignac}}, \bibinfo {author} {\bibfnamefont
  {R.}~\bibnamefont {Citro}}, \ and\ \bibinfo {author} {\bibfnamefont
  {T.}~\bibnamefont {Giamarchi}},\ }\href {\doibase
  10.1103/PhysRevLett.101.137207} {\bibfield  {journal} {\bibinfo  {journal}
  {Phys. Rev. Lett.}\ }\textbf {\bibinfo {volume} {101}},\ \bibinfo {pages}
  {137207} (\bibinfo {year} {2008})}\BibitemShut {NoStop}%
\bibitem [{\citenamefont {Tranquada}\ \emph {et~al.}(1995)\citenamefont
  {Tranquada}, \citenamefont {Sternlieb}, \citenamefont {Axe}, \citenamefont
  {Nakamura},\ and\ \citenamefont {Uchida}}]{tranquada1995}%
  \BibitemOpen
  \bibfield  {author} {\bibinfo {author} {\bibfnamefont {J.~M.}\ \bibnamefont
  {Tranquada}}, \bibinfo {author} {\bibfnamefont {B.~J.}\ \bibnamefont
  {Sternlieb}}, \bibinfo {author} {\bibfnamefont {J.~D.}\ \bibnamefont {Axe}},
  \bibinfo {author} {\bibfnamefont {Y.}~\bibnamefont {Nakamura}}, \ and\
  \bibinfo {author} {\bibfnamefont {S.}~\bibnamefont {Uchida}},\ }\href@noop {}
  {\bibfield  {journal} {\bibinfo  {journal} {Nature}\ }\textbf {\bibinfo
  {volume} {375}},\ \bibinfo {pages} {561} (\bibinfo {year}
  {1995})}\BibitemShut {NoStop}%
\bibitem [{\citenamefont {Omran}\ \emph {et~al.}(2015)\citenamefont {Omran},
  \citenamefont {Boll}, \citenamefont {Hilker}, \citenamefont {Kleinlein},
  \citenamefont {Salomon}, \citenamefont {Bloch},\ and\ \citenamefont
  {Gross}}]{omran2015}%
  \BibitemOpen
  \bibfield  {author} {\bibinfo {author} {\bibfnamefont {A.}~\bibnamefont
  {Omran}}, \bibinfo {author} {\bibfnamefont {M.}~\bibnamefont {Boll}},
  \bibinfo {author} {\bibfnamefont {T.~A.}\ \bibnamefont {Hilker}}, \bibinfo
  {author} {\bibfnamefont {K.}~\bibnamefont {Kleinlein}}, \bibinfo {author}
  {\bibfnamefont {G.}~\bibnamefont {Salomon}}, \bibinfo {author} {\bibfnamefont
  {I.}~\bibnamefont {Bloch}}, \ and\ \bibinfo {author} {\bibfnamefont
  {C.}~\bibnamefont {Gross}},\ }\href {\doibase 10.1103/PhysRevLett.115.263001}
  {\bibfield  {journal} {\bibinfo  {journal} {Phys. Rev. Lett.}\ }\textbf
  {\bibinfo {volume} {115}},\ \bibinfo {pages} {263001} (\bibinfo {year}
  {2015})}\BibitemShut {NoStop}%
\bibitem [{\citenamefont {Hilker}\ \emph {et~al.}(2017)\citenamefont {Hilker},
  \citenamefont {Salomon}, \citenamefont {Grusdt}, \citenamefont {Omran},
  \citenamefont {Boll}, \citenamefont {Demler}, \citenamefont {Bloch},\ and\
  \citenamefont {Gross}}]{hilker2017}%
  \BibitemOpen
  \bibfield  {author} {\bibinfo {author} {\bibfnamefont {T.~A.}\ \bibnamefont
  {Hilker}}, \bibinfo {author} {\bibfnamefont {G.}~\bibnamefont {Salomon}},
  \bibinfo {author} {\bibfnamefont {F.}~\bibnamefont {Grusdt}}, \bibinfo
  {author} {\bibfnamefont {A.}~\bibnamefont {Omran}}, \bibinfo {author}
  {\bibfnamefont {M.}~\bibnamefont {Boll}}, \bibinfo {author} {\bibfnamefont
  {E.}~\bibnamefont {Demler}}, \bibinfo {author} {\bibfnamefont
  {I.}~\bibnamefont {Bloch}}, \ and\ \bibinfo {author} {\bibfnamefont
  {C.}~\bibnamefont {Gross}},\ }\href {\doibase 10.1126/science.aam8990}
  {\bibfield  {journal} {\bibinfo  {journal} {Science}\ }\textbf {\bibinfo
  {volume} {357}},\ \bibinfo {pages} {484} (\bibinfo {year}
  {2017})}\BibitemShut {NoStop}%
\bibitem [{\citenamefont {Kruis}\ \emph {et~al.}(2004)\citenamefont {Kruis},
  \citenamefont {McCulloch}, \citenamefont {Nussinov},\ and\ \citenamefont
  {Zaanen}}]{kruis2004}%
  \BibitemOpen
  \bibfield  {author} {\bibinfo {author} {\bibfnamefont {H.~V.}\ \bibnamefont
  {Kruis}}, \bibinfo {author} {\bibfnamefont {I.~P.}\ \bibnamefont
  {McCulloch}}, \bibinfo {author} {\bibfnamefont {Z.}~\bibnamefont {Nussinov}},
  \ and\ \bibinfo {author} {\bibfnamefont {J.}~\bibnamefont {Zaanen}},\ }\href
  {\doibase 10.1103/PhysRevB.70.075109} {\bibfield  {journal} {\bibinfo
  {journal} {Phys. Rev. B}\ }\textbf {\bibinfo {volume} {70}},\ \bibinfo
  {pages} {075109} (\bibinfo {year} {2004})}\BibitemShut {NoStop}%
\bibitem [{\citenamefont {Ogata}\ and\ \citenamefont
  {Shiba}(1990)}]{ogata1990}%
  \BibitemOpen
  \bibfield  {author} {\bibinfo {author} {\bibfnamefont {M.}~\bibnamefont
  {Ogata}}\ and\ \bibinfo {author} {\bibfnamefont {H.}~\bibnamefont {Shiba}},\
  }\href {\doibase 10.1103/PhysRevB.41.2326} {\bibfield  {journal} {\bibinfo
  {journal} {Phys. Rev. B}\ }\textbf {\bibinfo {volume} {41}},\ \bibinfo
  {pages} {2326} (\bibinfo {year} {1990})}\BibitemShut {NoStop}%
\bibitem [{\citenamefont {Woynarovich}(1982)}]{woynarovich1982}%
  \BibitemOpen
  \bibfield  {author} {\bibinfo {author} {\bibfnamefont {F.}~\bibnamefont
  {Woynarovich}},\ }\href {\doibase 10.1088/0022-3719/15/1/007} {\bibfield
  {journal} {\bibinfo  {journal} {J. Phys. C Solid State}\ }\textbf {\bibinfo
  {volume} {15}},\ \bibinfo {pages} {85} (\bibinfo {year} {1982})}\BibitemShut
  {NoStop}%
\bibitem [{\citenamefont {Bogoliubov}\ \emph {et~al.}(1986)\citenamefont
  {Bogoliubov}, \citenamefont {Izergin},\ and\ \citenamefont
  {Korepin}}]{bogoliubov1986}%
  \BibitemOpen
  \bibfield  {author} {\bibinfo {author} {\bibfnamefont {N.}~\bibnamefont
  {Bogoliubov}}, \bibinfo {author} {\bibfnamefont {A.}~\bibnamefont {Izergin}},
  \ and\ \bibinfo {author} {\bibfnamefont {V.}~\bibnamefont {Korepin}},\ }\href
  {\doibase https://doi.org/10.1016/0550-3213(86)90579-1} {\bibfield  {journal}
  {\bibinfo  {journal} {Nuclear Physics B}\ }\textbf {\bibinfo {volume}
  {275}},\ \bibinfo {pages} {687} (\bibinfo {year} {1986})}\BibitemShut
  {NoStop}%
\bibitem [{\citenamefont {Brinkman}\ and\ \citenamefont
  {Rice}(1970)}]{brinkman1970}%
  \BibitemOpen
  \bibfield  {author} {\bibinfo {author} {\bibfnamefont {W.~F.}\ \bibnamefont
  {Brinkman}}\ and\ \bibinfo {author} {\bibfnamefont {T.~M.}\ \bibnamefont
  {Rice}},\ }\href {\doibase 10.1103/PhysRevB.2.1324} {\bibfield  {journal}
  {\bibinfo  {journal} {Phys. Rev. B}\ }\textbf {\bibinfo {volume} {2}},\
  \bibinfo {pages} {1324} (\bibinfo {year} {1970})}\BibitemShut {NoStop}%
\bibitem [{\citenamefont {Greif}\ \emph {et~al.}(2015)\citenamefont {Greif},
  \citenamefont {Jotzu}, \citenamefont {Messer}, \citenamefont {Desbuquois},\
  and\ \citenamefont {Esslinger}}]{greif2015}%
  \BibitemOpen
  \bibfield  {author} {\bibinfo {author} {\bibfnamefont {D.}~\bibnamefont
  {Greif}}, \bibinfo {author} {\bibfnamefont {G.}~\bibnamefont {Jotzu}},
  \bibinfo {author} {\bibfnamefont {M.}~\bibnamefont {Messer}}, \bibinfo
  {author} {\bibfnamefont {R.}~\bibnamefont {Desbuquois}}, \ and\ \bibinfo
  {author} {\bibfnamefont {T.}~\bibnamefont {Esslinger}},\ }\href {\doibase
  10.1103/PhysRevLett.115.260401} {\bibfield  {journal} {\bibinfo  {journal}
  {Phys. Rev. Lett.}\ }\textbf {\bibinfo {volume} {115}},\ \bibinfo {pages}
  {260401} (\bibinfo {year} {2015})}\BibitemShut {NoStop}%
\bibitem [{\citenamefont {White}\ and\ \citenamefont
  {Affleck}(2001)}]{white2001}%
  \BibitemOpen
  \bibfield  {author} {\bibinfo {author} {\bibfnamefont {S.~R.}\ \bibnamefont
  {White}}\ and\ \bibinfo {author} {\bibfnamefont {I.}~\bibnamefont
  {Affleck}},\ }\href {\doibase 10.1103/physrevb.64.024411} {\bibfield
  {journal} {\bibinfo  {journal} {Phys. Rev. B}\ }\textbf {\bibinfo {volume}
  {64}},\ \bibinfo {pages} {024411} (\bibinfo {year} {2001})}\BibitemShut
  {NoStop}%
\bibitem [{\citenamefont {Grusdt}\ \emph {et~al.}(2017)\citenamefont {Grusdt},
  \citenamefont {Kanasz-Nagy}, \citenamefont {Bohrdt}, \citenamefont {Chiu},
  \citenamefont {Ji}, \citenamefont {Greiner}, \citenamefont {Greif},\ and\
  \citenamefont {Demler}}]{grusdt2017}%
  \BibitemOpen
  \bibfield  {author} {\bibinfo {author} {\bibfnamefont {F.}~\bibnamefont
  {Grusdt}}, \bibinfo {author} {\bibfnamefont {M.}~\bibnamefont {Kanasz-Nagy}},
  \bibinfo {author} {\bibfnamefont {A.}~\bibnamefont {Bohrdt}}, \bibinfo
  {author} {\bibfnamefont {C.~S.}\ \bibnamefont {Chiu}}, \bibinfo {author}
  {\bibfnamefont {G.}~\bibnamefont {Ji}}, \bibinfo {author} {\bibfnamefont
  {M.}~\bibnamefont {Greiner}}, \bibinfo {author} {\bibfnamefont
  {D.}~\bibnamefont {Greif}}, \ and\ \bibinfo {author} {\bibfnamefont
  {E.}~\bibnamefont {Demler}},\ }\href@noop {} {\bibfield  {journal} {\bibinfo
  {journal} {arXiv:1712.01874}\ } (\bibinfo {year} {2017})}\BibitemShut
  {NoStop}%
\bibitem [{\citenamefont {Dagotto}\ and\ \citenamefont
  {Rice}(1996)}]{dagotto1996}%
  \BibitemOpen
  \bibfield  {author} {\bibinfo {author} {\bibfnamefont {E.}~\bibnamefont
  {Dagotto}}\ and\ \bibinfo {author} {\bibfnamefont {T.~M.}\ \bibnamefont
  {Rice}},\ }\href {\doibase 10.1126/science.271.5249.618} {\bibfield
  {journal} {\bibinfo  {journal} {Science}\ }\textbf {\bibinfo {volume}
  {271}},\ \bibinfo {pages} {618} (\bibinfo {year} {1996})}\BibitemShut
  {NoStop}%
\bibitem [{\citenamefont {White}\ and\ \citenamefont
  {Scalapino}(1997)}]{white1997}%
  \BibitemOpen
  \bibfield  {author} {\bibinfo {author} {\bibfnamefont {S.~R.}\ \bibnamefont
  {White}}\ and\ \bibinfo {author} {\bibfnamefont {D.~J.}\ \bibnamefont
  {Scalapino}},\ }\href {\doibase 10.1103/PhysRevB.55.6504} {\bibfield
  {journal} {\bibinfo  {journal} {Phys. Rev. B}\ }\textbf {\bibinfo {volume}
  {55}},\ \bibinfo {pages} {6504} (\bibinfo {year} {1997})}\BibitemShut
  {NoStop}%
\bibitem [{\citenamefont {White}\ and\ \citenamefont
  {Scalapino}(1998)}]{white1998}%
  \BibitemOpen
  \bibfield  {author} {\bibinfo {author} {\bibfnamefont {S.~R.}\ \bibnamefont
  {White}}\ and\ \bibinfo {author} {\bibfnamefont {D.~J.}\ \bibnamefont
  {Scalapino}},\ }\href {\doibase 10.1103/PhysRevLett.80.1272} {\bibfield
  {journal} {\bibinfo  {journal} {Phys. Rev. Lett.}\ }\textbf {\bibinfo
  {volume} {80}},\ \bibinfo {pages} {1272} (\bibinfo {year}
  {1998})}\BibitemShut {NoStop}%
\bibitem [{\citenamefont {B\"uchler}(2010)}]{buchler2010}%
  \BibitemOpen
  \bibfield  {author} {\bibinfo {author} {\bibfnamefont {H.~P.}\ \bibnamefont
  {B\"uchler}},\ }\href {\doibase 10.1103/PhysRevLett.104.090402} {\bibfield
  {journal} {\bibinfo  {journal} {Phys. Rev. Lett.}\ }\textbf {\bibinfo
  {volume} {104}},\ \bibinfo {pages} {090402} (\bibinfo {year}
  {2010})}\BibitemShut {NoStop}%
\bibitem [{\citenamefont {Jordan}\ and\ \citenamefont
  {Wigner}(1928)}]{jordan1928}%
  \BibitemOpen
  \bibfield  {author} {\bibinfo {author} {\bibfnamefont {P.}~\bibnamefont
  {Jordan}}\ and\ \bibinfo {author} {\bibfnamefont {E.}~\bibnamefont
  {Wigner}},\ }\href {\doibase 10.1007/BF01331938} {\bibfield  {journal}
  {\bibinfo  {journal} {Zeitschrift f{\"u}r Physik}\ }\textbf {\bibinfo
  {volume} {47}},\ \bibinfo {pages} {631} (\bibinfo {year} {1928})}\BibitemShut
  {NoStop}%
\bibitem [{\citenamefont {Prokof'ev}\ \emph {et~al.}(1998)\citenamefont
  {Prokof'ev}, \citenamefont {Svistunov},\ and\ \citenamefont
  {Tupitsyn}}]{prokofev1998}%
  \BibitemOpen
  \bibfield  {author} {\bibinfo {author} {\bibfnamefont {N.~V.}\ \bibnamefont
  {Prokof'ev}}, \bibinfo {author} {\bibfnamefont {B.~V.}\ \bibnamefont
  {Svistunov}}, \ and\ \bibinfo {author} {\bibfnamefont {I.~S.}\ \bibnamefont
  {Tupitsyn}},\ }\href {\doibase 10.1134/1.558661} {\bibfield  {journal}
  {\bibinfo  {journal} {Journal of Experimental and Theoretical Physics}\
  }\textbf {\bibinfo {volume} {87}},\ \bibinfo {pages} {310} (\bibinfo {year}
  {1998})}\BibitemShut {NoStop}%
\bibitem [{\citenamefont {Pollet}\ \emph {et~al.}(2007)\citenamefont {Pollet},
  \citenamefont {Houcke},\ and\ \citenamefont {Rombouts}}]{pollet2007}%
  \BibitemOpen
  \bibfield  {author} {\bibinfo {author} {\bibfnamefont {L.}~\bibnamefont
  {Pollet}}, \bibinfo {author} {\bibfnamefont {K.~V.}\ \bibnamefont {Houcke}},
  \ and\ \bibinfo {author} {\bibfnamefont {S.~M.}\ \bibnamefont {Rombouts}},\
  }\href {\doibase https://doi.org/10.1016/j.jcp.2007.03.013} {\bibfield
  {journal} {\bibinfo  {journal} {Journal of Computational Physics}\ }\textbf
  {\bibinfo {volume} {225}},\ \bibinfo {pages} {2249 } (\bibinfo {year}
  {2007})}\BibitemShut {NoStop}%
\end{thebibliography}%

\end{document}